\documentstyle[12pt]{article}                  

\begin{document}

\begin{titlepage}
\title{ON THE CONSTRUCTION OF \\ RENORMALIZED GAUGE THEORIES \\
USING RENORMALIZATION GROUP TECHNIQUES }

\author{C. BECCHI\\
Dipartimento di Fisica, Universit\`a di Genova,\\
Istituto Nazionale di Fisica Nucleare, Sezione di Genova,\\
via Dodecaneso 33, 16146 Genova (Italy)}
\date{}

\maketitle

\begin{abstract}
The aim of these lectures is to describe a construction, as self-contained as 
possible, of renormalized gauge theories.
Following a suggestion of Polchinski, we base our analysis
on the Wilson renormalization group method. 

After a discussion of the infinite cut-off limit, we study the 
short distance properties of the Green functions verifying 
the validity of Wilson short
distance expansion. We also consider the problem of  the extension to the 
quantum level of the classical symmetries of the theory. With this purpose 
we analyze in details the breakings induced by the 
cut-off in a $SU(2)$ gauge symmetry and we prove the possibility of 
compensating these breakings by a suitable choice of non-gauge invariant 
counter terms.
\end{abstract}
\vfill \footnote{Lectures given at
the Parma Theoretical Physics Seminar, September 1991 - Revised Version}

\eject

\end{titlepage}

\renewcommand{\theequation}{\thesection ,\arabic{equation}}
\newcommand{\newsection}[1]{
\vspace{15mm}\pagebreak[3]
\addtocounter{section}{1}
\setcounter{equation}{0}
\begin{flushleft}
{\large\bf \thesection. #1}
\end{flushleft}
\nopagebreak
\medskip
\nopagebreak}

\def\sec{\newsection}
\def\ss{\subsection}
\def \cp{\chapter}
\def\be{\begin{equation}}
\def\ee{\end{equation}}
\def\bea{\begin{eqnarray}&&}
\def\eea{\end{eqnarray}}
\def\nn{\nonumber \\ &&}
\def\acca{\right.\nn\left.}

\def\sp{\par\vfill\eject}
\def\salto{\par\vskip .5cm}
\def\quasisalto{\par\vskip .4cm}
\def\saltone{\par\vskip .7cm}
\def\saltino{\par\vskip .2cm}
\def\capo{\par\noindent}

\def\t{\tilde}
\def\D{\Delta}
\def\De{{\Delta_{eff}}}
\def\Dl{{{\cal T}\Delta_{eff}}}
\def\hDe{{D_{eff}}}
\def\hDl{{{\cal T}D_{eff}}}
\def\o{\over}
\def\g{\Gamma}

\def\i{\int d^4x}
\def\p{\partial}
\def\ep{\epsilon^{a b c}}
\def\a{\alpha}
\def\b{\beta}
\def\c{\gamma}
\def\d{\delta}
\def\i2{\int d^2x}
\def\v{\vec}
\def\k{K_{\Lambda}\phi}
\def\L{L_{eff}}
\def\S{{S_{eff}}}
\def\Sl{{{\cal T}S_{eff}}}
\def\f{\phi}
\def\la{\Lambda}
\def\lao{\Lambda_0}
\def\ko{K_{\Lambda_0}\phi}
\def\in{\int\prod_{\vec p}d\phi_{\vec p}}
\def\u{\omega}
\def\({\left(}
\def\){\right)}
\def\[{\left[}
\def\]{\right]}
\def\lf{{\cal L}_{eff}}
\def\ll{l_{eff}}
\def\ipp{\int d^4p}
\def\ippp{\ipp {\D '(p)\o (2\pi)^4}}
\def\df{\d \o\d\f (}
\def\vp{\vec p}
\def\vq{\vec q}
\def\da{\partial_{A_{\mu -\vec p}^a}}
\def\dc{\partial_{c_{ -\vec p}^a}}
\def\dg{\partial_{\gamma_{\mu \vec p}^a}}
\def\dz{\partial_{\zeta_{\vec p}^a}}
\def\dea{{\d\o\d\t A_{\mu }^a(p)}}
\def\dec{{\d\o\d\t c ( p)^a}}
\def\deg{{\d\o\d\t \gamma_{\mu}^a(-p)}}
\def\dez{{\d\o\d\t \zeta (-p)^a}}
\def\SP{\sum_{\vec p}}
\def\SQ{\sum_{\vec q}}
\def\r.{\right.}
\def\l.{\left.}
\def\IP{\int d^4p}
\def\pp{\p_{\vp}}
\def\pmp{\p_{-\vp}}
\def\pq{\p_{\vq}}
\def\pmq{\p_{-\vq}}
\def\pgp{\p_{\c_{\vp}}}
\def\pgmp{\p_{\c_{-\vp}}}
\def\pgq{\p_{\c_{\vq}}}
\def\pgmq{\p_{\c_{-\vq}}}
\def\fp{\f_{\vp}}
\def\fq{\f_{\vq}}
\def\ldl{\la\p_{\la}}
\def\ldo{\lao\p_{\lao}}
\def\kpo{k\({p\o\Lambda_0}\)}
\def\kq{k\({q\o\Lambda}\)}
\def\kp{k\({p\o\Lambda}\)}
\def\kdq{k^2\({q\o\Lambda}\)}
\def\kdp{k^2\({p\o\Lambda}\)}

\salto\noindent
\sec{Introduction}
\noindent\salto

In a period of approximately ten years, about twenty years ago,
the perturbative construction of renormalized quantum field theory has achieved
a remarkable level of rigor and efficiency.
based on many technical achievements that are widely 
explained in the current literature. We mention among others the extension to all 
orders of perturbation theory of subtraction schemes suitable to avoid
infinities in the Feynman amplitudes. This has put on a rigorous basis the
method of counter-terms, although in the framework of a formal perturbation
theory which is not absolutely summable.

A second significant technical progress consists in the discovery of very
clever regularization schemes, in particular the dimensional one
which has made possible calculations of renormalization effects of 
remarkably high order, as e. g. the fourth of QCD. This has greatly 
improved the efficiency of renormalized quantum field theory .

The great majority of the above mentioned achievements have been based
on a deep and complicated analysis of Feynman diagrams a typical ingredient
of which is the concept of "forest". The difficulties with 
Feynman diagrams are amplified in the gauge models of fundamental 
interactions in which the number of contributions to a given amplitude
increases rapidly with the perturbative order and hence it is often
prohibitive to push the calculations beyond one loop. In these models
dimensional regularization too can become a source of problems due to the 
difficulty of extending the concept of chirality to 
complex space-time dimensions.

Waiting for new ideas and tools to compute higher orders in gauge theories,
there remains, in our opinion, the need of an essential simplification of the
proofs of the relevant general properties of renormalized quantum field theory.

Few years ago Polchinski \cite{1} has shown how the existence of the
ultra-violet limit of a scalar theory, regularized by means of a momentum 
cut-off, can be proved using Wilson renormalization group techniques 
\cite{2} .
The method of Polchinski is remarkably simple and can be trivially extended
to spinor and vector fields; however one still needs to recover in the same
framework the whole set of general results that have made possible
the above mentioned progresses in renormalized quantum field theory.

In these lectures we present an attempt, following
the ideas of Polchinski, to give a self-contained proof
of the existence of a perturbatively
renormalized quantum field theory and of two "general properties" 
of it. First the validity of Wilson short distance expansion \cite{3} 
which is a 
fundamental tool in the analysis of Green functions. Secondly we discuss
the "quantum action principle", from which rather general results on the
renormalized structure of theories with continuous symmetries can be 
obtained \cite{4},\cite{5} .

The lecture notes are so organized: in section 2 we discuss the construction 
of the Feynman functional generator. In section 3 we present the renormalization
group method, whose perturbative solution is discussed in section 4.
In section 5 we make some comments on the construction of composite 
operators and we discuss Wilson short distance expansions. In section 6
we derive the "quantum action principle" that we apply to the study of the 
$SU(2)$ Yang-Mills model.
\vfill\eject
\salto
\sec{The Feynman formula}
\salto

The existence of a theory of scattering is one of the fundamental results
of field theory. It is based on the classical axiomatic results on the
asymptotic evolution of states and on the well known 
Lehmann-Symanzik-Zimmermann reduction formulae 
relating  $S$ matrix elements to time-ordered Green 
functions. If $\phi_{in}$ is a set of asymptotic ingoing
fields whose wave operator is $W$ we can write the scattering 
operator $S_{op}$ in the
Fock space of $\phi_{in}$ as:
\be  S_{op}=:exp\( \int d^4x \f_{in} (x) W_x z^{-1} {\d\o\d j(x)}\):Z[j]\vert_{j=0}
\ ,\label{2,1}\ee
where $z$ is the residue of the Fourier transformed two-point function
on the mass shell pole and the functional $Z[j]$ is the generator of the 
time-ordered Green functions:
\be  Z[j]=(\Omega ,Te^{i\int d^4x j(x)\phi (x)}\Omega)\ ,\label{2,2}\ee
and $\Omega$ is the vacuum state.

The determination of the functional generator 
$Z$ is therefore the main dynamical
problem in the construction of a field theory. The Feynman formula is 
universally considered as the solution of this problem.

This formula is usually written i terms of the classical action $S_{cl}$ according:
\be Z[j]=N\int\prod_x d\phi (x)e^{i\[S_{cl}-\int d^4x j(x)\phi (x)\]}\ 
,\label{2,6}\ee where it is assumed the functional integral to make sense and 
the measure $\prod_x d\phi (x)$ to be translation invariant.

In the first part of this course we shall discuss the possibility giving a 
meaning to this formula. For this we have to overcome a sequence of 
difficulties that can be traced back to nature of 
the functional measure and to that of the 
integrand. 

The problem with the measure is technically 
related to the lack of local compactness, that is, to the presence 
of an infinite number of variables. Concerning the integrand, if the 
measure is translation invariant as assumed, it has to introduce a uniform 
convergence factor for large field amplitudes. This is not verified in the 
present case since the integrand has absolute value equal to one.

The standard solution to this convergence problem is based on the 
transformation of the minkowskian theory into an euclidean one.  
This is achieved
"Wick rotating" the time variables from the positive real axis to the negative 
imaginary one.
The 
time-ordered Green functions are then replaced by the euclidean Schwinger 
functions \cite{6} whose functional generator is defined
in complete analogy with (\ref{2,6}), absorbing the imaginary 
unit into the euclidean space measure. That is:
\be Z[j]=N\int\prod_x d\phi (x)e^{-\[S_{e}-\int d^4x j(x)\phi (x)\]}\ 
.\label{2,2e}\ee 
If the euclidean classical action 
$S_e$ is a positive functional increasing with the field 
amplitude the wanted convergence factor is guaranteed.
Of course, we have overcome the first difficulty at the price of computing 
something that is different from our goal. However, Osterwalder and 
Schrader have identified the conditions ensuring the possibility of 
recovering the wanted physical information from an euclidean 
theory. \cite{h}

Coming back to the number of integration variables,
we notice that one can distinguish two sources of this difficulty.
First, euclidean invariance requires the space volume to be infinite. This 
is the infra-red (IR) difficulty. 
Giving up the euclidean invariance 
one could quantize the theory in a hypercube choosing as integration 
variables the Fourier 
amplitudes of the field. However these are still infinite in number and 
locality requires the interaction to involve all the field Fourier 
components. This is the ultra-violet (UV) difficulty, it can also be seen from
 a different point of view. If the interacting Green (Schwinger) functions
are distributions, as in the free case, the same is true for the 
functional derivatives of $Z$ and hence the interaction cannot be written
as a strictly local functional of the fields.

To avoid these IR and UV difficulties, giving up for the moment locality 
and covariance, we introduce a system of regularizations. First, as said 
above, 
we restrict our theory into a four-dimensional hypercube $\Omega$.
 Periodic boundary 
conditions are chosen for the fields in order to preserve translation 
invariance. As a matter of fact in this way we are considering a quantum 
mechanical system in a three-dimensional cube of side $L$ in thermal
equilibrium at the temperature $\beta^{-1}={1\o L}$. An infinite volume and
zero temperature limit will eventually reproduce the original relativistic
euclidean field theory.

The second regularization concerns directly the UV difficulties.
We modify the interaction decoupling the short wave length modes which
, however, continue to appear into the dynamical framework as free 
degrees of freedom. Notice that in standard approaches these degrees of freedom
are simply not taken into account.

We regularize the euclidean theory by
replacing into the interaction the field $\f (x)$ with
\be \ko (x)={1\o L^2}\sum_{\v p} k({p^2\o \lao^2})\f_{\v p} e^{ipx}\  
\label{2,16}\ee
where \be \f_{\v p}={1\o L^2}\int_{\Omega}
 d^4 x\f (x)e^{ipx}\  ,\label{2,17}\ee and, 
of course the integral
is limited within the above mentioned four-dimensional hypercube of side $L$.
The ultraviolet (UV) cut-off factor
$k$ is a $C^{\infty}$ function assuming the value 1 below 1 and
vanishing above 2.

It should be clear that the introduction of the UV cut-off should interfere
as weakly as possible with the observables of our theory. This requires, in 
particular, that the chosen value of the cut-off be much higher than the 
greatest wave number appearing in the Fourier decomposition of the sources
$j$ and of the other sources that one could introduce to define composite 
operators. 
More precisely let $\la_R$ be the greatest observable wave number, we 
select the sources so that \be K_{\la_R} j=j\ .\label{2,18}\ee 
and we ask
\be \la_0\gg\la_R\ .\label{2,20}\ee
Eventually $\lao$ will be sent to infinity.

Now we come to the choice of the action. This is done having in 
mind the short-distance properties of the free theory that we want only 
weakly perturbed by the interaction. With this purpose we assign
to every field $\f$ a mass-dimension $d_{\f}$ and to 
the corresponding source $j$, ($d_j=D-d_{\f}$), 
where $D$ is the dimension of the euclidean
space (in our case 4). $d_{\f}$ and hence $d_j$ are computed from the 
mass-dimension of the wave operator $W$ requiring that the free field equations
be dimensionally homogeneous. Thus, in the scalar field case, the 
mass-dimension of the wave operator is two, that of the laplacian, and hence
the mass-dimension of the field is computed from:
\be 2+d_{\f}=d_j=D-d_{\f}\ .\label{2,10}\ee 
That is one in the 4-dimensional case.

The action in (\ref{2,2e}) is chosen as the sum of the free part
\be \int_{\Omega} d^4x{\(\p \f\)^2+m^2\f^2\o2}\ ,\ee and of the interaction 
$L_0$, an integrated local 
polynomial in the regularized field $\k$ and its 
derivatives. The short-distance "power counting" condition
limits to 4 the dimension of the operators appearing in $L_0$.
Therefore in the scalar case we have:
\be L_0=\int_{\Omega} d^4x \[{\lambda_0\o 4!}\f^4+{z_0-1\o 2}(\p\f)^2+{m_0^2-m^2\o 2}\f^2
\]\ .\label{2,8}\ee
A particular care should be devoted to the definition of composite operators.
To introduce for example the operator $\f^2$ we assign him the source 
$\u$, whose dimension is the complement to 4 of the dimension of the 
operator, that is 2, and we add to the interaction $L_0$ the $\u$-dependent 
terms:
\be \int_{\Omega} d^4x \[\zeta_0\u\f^2+\eta_0 \u^2(x)+\xi_0\u(x)\]\ .\label{2,11}\ee
This functional identifies the general solution of 
an extended power counting condition 
taking into account also the dimension of the sources.

 One should wonder about the meaning 
of the last two terms that are field independent. It is clear from the definition of
the functional generator $Z$ that e. g. the $\u^2$ term induces a contribution 
proportional to a Dirac $\d$ in the correlation function of two composite
operators. Back to the minkowskian world this corresponds to a
redefinition of the time-ordering of the product of two operators.

Now, taking into account both UV and IR regularizations, we write the 
interaction of our model according:
\bea L_0=\int_{\Omega} d^4x 
 \[{\lambda_0\o 4!}(\ko)^4+{z_0-1\o 2}(\p\ko)^2+
{m_0^2-m^2\o 2}(\ko)^2 +\right.\nn \left.
\zeta_0\u(\ko)^2 + \eta_0 \u^2+\xi_0\u\]= 
{\lambda_0\o 4!L^4}\sum_{\v p_1,..\v p_4}
\d_{\v 0,\v p_1+..+\v p_4}\,\ko_{\v p_1}...\ko_{\v p_4}+\nn  
 \sum_{\v p} {(z_o-1)p^2+m_0^2-m^2\o 2}\ko_{\v p}\ko_{-\v p}+ \nn
{\zeta_0\o L^2}\sum_{\v p_1,..\v p_3}
\d_{\v 0,\v p_1+..+\v p_3}\,\u_{\v p_1}\ko_{\v p_2}\ko_{-\v p_3}+\nn
\eta_0\sum_{\v p} \u_{\v p}\u_{-\v p}+
L^2\xi_0\u_{\v 0}\ .\label{2,21}\eea
Notice that by the IR regularization we have automatically broken the 
euclidean invariance of the theory, since a cube is not rotation invariant.
Therefore there is no reason to preserve the euclidean invariance of every 
single term of the 
interaction. Following the fine-tuning strategy which will be discussed in 
section 6 it is possible to prove the compensability of the 
possible breakdown of euclidean invariance induced by the IR regularization
 by the introduction into the 
interaction of non-invariant counter-terms. 
This compensability holds true for all continuous
symmetries if the symmetry group is semisimple. This
is a typical consequence of the quantum action principle \cite{4},\cite{5}.

It is apparent that in our example $L_0$ depends on the 6 parameters
$\lambda_0$, $z_0$, $m_0$, $\zeta_0$, $\eta_0$ and $\xi_0$ that can be 
extracted from the interaction by suitable normalization operators. That is:
\be \rho_{0,1}=L^4{\p^4\o\p\f_{\v 0}^4} L_0\vert_{\f=\u=0}
\equiv N_1L_0=\lambda_0\ ,\label{2,22}\ee
and, setting
\be {\p\o\p\f_{\v p}}{\p\o\p\f_{-\v p}}L_0\vert_{\f=\u=0}=\Pi_{\v p}\
 ,\label{2,23}\ee
\be \rho_{0,2}={\Pi_{\v p}-\Pi_{\v 0}\o p^2}\equiv N_2L_0=z_0-1\ ,\label{
2,24}\ee 
for some 
suitably chosen $\v p$ and
\be \rho_{0,3}=\Pi_{\v 0}\equiv N_3L_0=m^2_0-m^2\ ,\label{2,25}\ee
\be \rho_{0,4}={L^2\o 2}{\p^3\o\p\f_{\v 0}^2\p\u_{\v 0}} L_0\vert_{\f=\u=0}
\equiv N_4 L_0=\zeta_0\ ,\label{2,26}\ee
\be \rho_{0,5}={1\o2}{\p^2\o\p\u_{\v 0}^2} 
L_0\vert_{\f=\u=0}\equiv N_5 L_0=\eta_0
\ ,\label{2,27}\ee
\be \rho_{0,6}=L^{-2}{\p\o\p\u_{\v 0}} L_0\vert_{\f=\u=0}\equiv N_6 L_0=\xi_0
\ .\label{2,28}\ee
The functional generator corresponding to the interaction (\ref{2,21}) is 
given by
\be Z[j,\u]=N\in e^{-\[\sum_{\v p}\f_{-\v p}{C(p)\o 2}\f_{\v p}+L_0-
\sum_{\v p}j_{-\v p}\f_{\v p}\]}\equiv N\in\  e^{-S}\ ,\label{2,30}\ee
where the normalization factor $N$ ensures the normalization condition
\be Z[0,0]=1\ .\label{2,31}\ee
and:
\be C(p)=p^2+m^2\ .\ee

Notice that, taking into account the 
 regularizations, the integral in (\ref{2,30})
factorizes in an infinite dimensional, purely gaussian, contribution
corresponding to the Fourier components $\f_{\v p}$ with $\v p > \sqrt 2\lao$,
that is reabsorbed into the normalization factor $N$, and in a finite
dimensional part, that is absolutely convergent provided that $\lambda_0$
is positive.

The crucial problem in quantum field theory is to prove the existence of an 
IR ($L\to\infty$) and an UV
($\lao\to\infty$) limit of (\ref{2,30}). This goal has been achieved in the 
framework of the perturbative method
for the whole family of theories satisfying the power counting criterion.

The major aim of these lectures is to describe the main lines of this 
result. For this is convenient to remember that
perturbation theory is constructed introducing into the Feynman functional 
integral the ordering parameter $\hbar$ according
\be Z[j,\u]\equiv e^{Z_c[j,\u]\o\hbar}= N\in\  e^{-S\o\hbar}\ ,\label{2,32}\ee
and developing the connected functional $Z_c$ as a formal power series in 
$\hbar$. This power series is obtained applying to (\ref{2,32}) the method of 
the steepest descent. Analyzing the terms of the series as  sums of  
Feynman diagrams one sees that $Z_c$ receives contributions only from the 
connected diagrams and that $\hbar$ can be interpreted as a loop counting 
parameter.   
\vfill\eject
\salto
\sec{The Wilson renormalization group method.}
\salto

 A fundamental tool in the analysis of the UV limit
is the Wilson renormalization group method. \cite{2}
This method, in its original 
version, is based on a sharp UV cut-off ($\lao$), or equivalently 
on a lattice regularization, limiting the number of degrees of freedom, i. e. 
that of the integration variables in the Feynman formula. It consists in the 
iterative reduction of the number of degrees of freedom through the integration 
over the field Fourier components $\f_{\v p}$ with $\lao\geq p\geq {\lao\o2}.$
 The resulting, partially integrated, Feynman formula 
is brought back to the original form substituting the interaction lagrangian
with an effective one which now depends on the fields cut-off at ${\lao\o2}$.
The existence of the wanted UV limit is related to that of a "fixed point"
in the space of effective lagrangians which is approached afteranalyzing a 
great number of iterations of the partial integration procedure.

Following Polchinski, \cite{1}
 we shall use a modified version of this 
method based on a continuous lowering of the cut-off, which, in our 
case, corresponds to a continuous switching-off of the interaction of the field
components with higher wave number.

This is accompanied by a continuous evolution of the effective 
lagrangian $\L$ replacing the bare 
interaction $L_0$ in the Feynman formula as it is 
exhibited in :
\bea Z[j,\u]=N\in e^{-\[\sum_{\v p}\f_{-\v p}{C(p)\o 2}\f_{\v p}+
\L(\k,\u,\la,\lao,\rho_0)
-\sum_{\v p}j_{-\v p}\f_{\v p}\]}\equiv 
\nn N\in\ e^{-S}\ ,\label{3,1}\eea
which holds true for $\la_R <\la <\lao$. The identity of (\ref{3,1}) and 
(\ref{2,30}) is guaranteed if $\L$ satisfies 
the following evolution equation:
\be \la\p_{\la}\L={1\o2}\sum_{\v p}\la\p_{\la}k^2\({p\o\la}\)C^{-1}(p)
\[\p_{-\v p}\L\p_{\v p}\L-\p_{\v p}\p_{-\v p}\L\]\ ,\label{3,2}\ee
with the initial condition \be \L(\ko,\u,\lao,\lao,\rho_0)=L_0\ .\label{
3,3}\ee
In (\ref{3,2}) we have introduced the simplified notation:
\be \p_{\v p}\L={\p\L\o\p\k_{\v p}}\ .\label{3,4}\ee

Using (\ref{3,2}) it is easy to verify (\ref{3,1}). Indeed
\be \la\p_{\la}Z[j,\u]=-\in\[\sum_{\v p}\la\p_{\la}\k_{\v p}\p_{\v p}\L+
\la\p_{\la}\L\]e^{-S}\ ,\label{3,5}\ee
and
\be e^S\ C^{-1}(p)\p_{\f_{-\v p}}e^{-S}=-\f_{\v p}-C^{-1}(p)\[
k\({p\o\la}\)\p_{-\v p}\L-j_{\v p}\]\ .\label{3,6}\ee
Thus we have:
\bea \la\p_{\la}Z[j,\u]=  -\in\[\la\p_{\la}\L-\right.\nn\left.
\sum_{\v p}\la\p_{\la}
k\({p\o\la}\)\p_{\v p}\L C^{-1}(p)\[\p_{\f_{-\v p}}+
k\({p\o\la}\)\p_{-\v p}\L
\]\] e^{-S}\ ,\label{3,7}\eea
where we have taken into account that from (\ref{2,18}) and (\ref{2,20}) one has
\be \la\p_{\la}k\({p\o\la}\)j_{\v p}=0\ .\label{3,8}\ee
Now, integrating by parts the second term in the right-hand side of 
(3,7) we get:
\bea \la\p_{\la}Z[j,\u]=-\in\[\la\p_{\la}\L-\right.\nn 
\left.\sum_{\v p}\la\p_{\la}k\({p\o\la}\)
C^{-1}(p)k\({p\o\la}\)
\[\p_{-\v p}\L\p_{\v p}\L-\p_{\v p}\p_{-\v p}\L\]\] e^{-S}\ 
.\label{3,9}\eea
This vanishes owing to (\ref{3,2}).

The evolution equation (\ref{3,2}) defines a family of lines, identified by the 
parameters $\rho_{0,a}$ of the bare lagrangian for a certain choice of the 
cut-off $\lao$,
in the space of the field functionals $\L$. The running parameter along the
lines is the cut-off $\la$. However these lines do not describe the evolution
toward the limit we are interested in. In order to prove the existence
of an UV limit of our Green functional $Z$, we should rather study the evolution
of our theory when $\lao$ increases and some "low energy" properties of
the effective interaction are kept fixed. We can define these low energy 
properties by means of the six numbers:
\be \rho_a(\la,\lao,\rho_0)=N_a\L\ ,\label{3,10}\ee
where $N_a$ ($a=1,..,6$) are the normalization operators defined in 
(\ref{2,22})-(\ref{2,28}).
These effective parameters are in one-to-one correspondence with the parameters 
of the bare interaction and we assume that the functional relation between
$\rho_{0,a}$ and $\rho_a$ be invertible. This will certainly be true at 
least for a limited range of $\la$. We denote by
\be \rho_{0,a}(\la,\lao,\rho)\label{3,11}\ee the inverse function of 
(\ref{3,10}).
Let us now consider the new field functional:
\bea 
V(\la)\equiv\lao\p_{\lao}\L\(\k,\u,\la,\lao,\rho_0(\la,\lao,\rho)\)=\nn
\lao\p_{\lao}\L-\lao\p_{\lao}\rho_a\p_{\rho_a}\L=\nn  
\lao\p_{\lao}\L(\k,\u,\la,\lao,\rho_0)-\nn
\lao\p_{\lao}\rho_a\[\({\p\rho\o\p\rho_0}\)^{-1}\]_b^a
\p_{\rho_{0,b}}\L(\k,\u,\la,\lao,\rho_0)\ .\label{3,12}\eea
This functional gives an indication of the dependence of $\L$ on the UV cut-off
when the parameters $\rho$ are kept fixed. We would like to show that 
$V$ vanishes stronger than a positive power of ${\la\o\lao}$. 
This would clearly indicate
that, once the low energy parameters are fixed, the whole theory tends 
toward a fixed point depending only on $\rho$. This is what is usually 
meant by renormalizability of the theory.

Let us then study the evolution equation of $V$. Given any field functional 
$X$, we define:\be M[X]\equiv\sum_{\v p}\la\p_{\la}k\({p\o\la}\)
C^{-1}(p)k\({p\o\la}\)
\[2\p_{-\v p}X\p_{\v p}\L-\p_{\v p}\p_{-\v p}X\]\ ,\label{3,13}\ee 
where the action of
$\p_{\v p}$ on $X$ is defined in the same way as on $\L$.

From (\ref{3,2}) we have:\be \la\p_{\la}\lao\p_{\lao}\L=M\[\lao\p_{\lao}\L\]\ 
.\label{3,14}\ee
This gives the evolution equation of the first term of $V$ in (\ref{3,12}).
 Considering
the second term, on gets three contributions which can be computed taking
into account that, by definition:
\be \la\p_{\la}\rho_a=N_a\[\la\p_{\la}\L\]\ .\label{3,15}\ee 
The first contribution is
\bea \la\p_{\la}\lao\p_{\lao}\rho_a\[\({\p\rho\o\p\rho_0}\)^{-1}\]_b^a
\p_{\rho_{0,b}}\L=\nn  N_a\[M\[\lao\p_{\lao}\L\]\]
\[\({\p\rho\o\p\rho_0}\)^{-1}\]_b^a
\p_{\rho_{0,b}}\L=\nn N_a\[M\[\lao\p_{\lao}\L\]\]\p_{\rho_a}\L\ 
.\label{3,16}\eea
The second is:\bea \lao\p_{\lao}\rho_a
\la\p_{\la}\[\({\p\rho\o\p\rho_0}\)^{-1}\]_b^a
\p_{\rho_{0,b}}\L=\nn  -
\lao\p_{\lao}\rho_a\[\({\p\rho\o\p\rho_0}\)^{-1}\]_c^a
\la\p_{\la}{\p\rho_d\o\p\rho_{0,c}}\[\({\p\rho\o\p\rho_0}\)^{-1}\]_b^d
\p_{\rho_{0,b}}\L=\nn  -
\lao\p_{\lao}\rho_a\[\({\p\rho\o\p\rho_0}\)^{-1}\]_c^a
N_d\[M\[{\p\L\o\p\rho_{0,c}}\]\]\[\({\p\rho\o\p\rho_0}\)^{-1}\]_b^d
\p_{\rho_{0,b}}\L\ .\label{3,17}\eea
The third is:
\bea \lao\p_{\lao}\rho_a
\[\({\p\rho\o\p\rho_0}\)^{-1}\]_b^a
\la\p_{\la}\p_{\rho_{0,b}}\L=\nn  \lao\p_{\lao}\rho_a
\[\({\p\rho\o\p\rho_0}\)^{-1}\]_b^aM\[{\p\L\o\p\rho_{0,b}}\]\ 
.\label{3,18}\eea
Combining (\ref{3,14}) and (\ref{3,18}) and recalling that $M$ is 
a linear operator we get $M[V]$.
In much the same way (\ref{3,14}) and (\ref{3,15}) give 
$N_a\[M[V]\]\p_{\rho_a}\L$.
Altogether we get:
\be \la\p_{\la}V=M[V]-N_a\[M[V]\]\p_{\rho_a}\L\ .\label{3,19}\ee
In order to compute the initial condition to this evolution equation we notice 
that, when $\la$ tends to $\lao$, $\L$ tends to $L_0$ and that
\be \rho(\lao,\lao,\rho_0)=\rho_0\ .\label{3,20}\ee Thus:
\bea \L\(\ko,\u,\la,\lao,\rho_0\)=L_0\(\ko,\u,\rho_0\)+ \nn  
(\la-\lao)\p_{\la}\L\(\ko,\u,\la,\lao,\rho_0\)\vert_{\la=\lao}+
O\((\la-\lao)^2\)\ .\label{3,21}\eea 
Then, taking the derivative of both members of (\ref{3,21}) 
with respect to $\lao$
and setting $\la=\lao$ we have:
\bea V[\lao]=\lao\p_{\lao} \L\(\k,\u,\la,\lao,\rho_0\) \vert_{\la=\lao}-\nn  
\lao\p_{\lao}\rho_a  \p_{\rho_a} \L\(\k,\u,\la,\lao,\rho_0 (\la ,\lao ,\rho\) 
 \vert_{\la=\lao}
=\nn  -\la\p_{\la}\L\(\ko,\u,\la,\lao,\rho_0\)\vert_{\la=\lao}+
\nn\la\p_{\la}N_a\[\L\(\ko,\u,\la,\lao,\rho_0\)\]\vert_{\la=\lao}\p_{\rho_a}L_0
=\nn  -{1\o2}\sum_{\v p}\lao\p_{\lao}k^2\(\({p\o\lao}\)^2\)
C^{-1}(p)\p_{-\v p}L_0\p_{\v p}L_0+\nn 
{1\o2}N_a\[\sum_{\v p}\lao\p_{\lao}k^2\(\({p\o\lao}\)^2\)
C^{-1}(p)\p_{-\v p}L_0\p_{\v p}L_0\]\p_{\rho_{0,a}}L_0+c\ ,\label{3,22}\eea 
where $c$ is field independent.

In deducing (\ref{3,22}) we have taken into account that, 
for any integrated local
functional $X$ of dimension up to 4, one has:
\be X=N_a\[X\]\p_{\rho_{0,a}}L_0+c\ ,\label{3,23}\ee where $c$ is again field 
independent.
We still need the evolution equation of
\be \[\({\p\rho\o\p\rho_0}\)^{-1}\]_b^a
\p_{\rho_{0,b}}\L\equiv \p_{\rho_a}\L\ .\label{3,24}\ee We have
\bea \la\p_{\la}\p_{\rho_a}\L=\[\({\p\rho\o\p\rho_0}\)^{-1}\]_b^a
\la\p_{\la}\p_{\rho_{0,b}}\L-\nn  
\[\({\p\rho\o\p\rho_0}\)^{-1}\]_c^a
\la\p_{\la}{\p\rho_d\o\p\rho_{0,c}}
 \[\({\p\rho\o\p\rho_0}\)^{-1}\]_b^d\p_{\rho_{0,b}}\L=\nn
 M\[\p_{\rho_a}\L\]-N_b\[M\[\p_{\rho_a}\L\]\]\p_{\rho_b}\L\ 
,\label{3,25}\eea
and the 
initial condition is:
\be \p_{\rho_a}\L\vert_{\la=\lao}=\p_{\rho_{0,a}}L_0\ .\label{3,26}\ee
It remains to discuss the solutions of the system of differential equations
(\ref{3,2}), (\ref{3,25}) and (\ref{3,19}) 
whose corresponding initial conditions 
are $L_0$, (\ref{3,26}) and (\ref{3,22}).

A first step in this discussion consists in the infinite volume limit. 
This is harmless at the level of the effective lagrangian since
the presence of a derivative of the cut-off function limits
the range of momenta in (\ref{3,2}) and (\ref{3,13}) between $\la$ and
$2\la$. This
and the euclidean character of our theory exclude any IR singularity.

We define the reduced functional $\hat\L$ according:
\be \L(\k,\u,\la,\lao,\rho_0)=(L\la )^4
\hat\L\({\k\o\la L^2},{\u\o\la^2 L^2},\la,\lao,\rho_0\)\ .\label{3,27}\ee
Writing $\L$ and $\hat\L$ as (formal) power series in $\k$ and $\u$ whose
coefficients, the effective vertices, 
are $\Gamma^{(n,m)}$ and $\hat\Gamma^{(n,m)}$. That is, for $\L$:
\bea \L=\sum^{\infty}_{n,m=0}{1\o n! m!}
\sum_{\v p_1,..,\v p_n,\v q_1,..,\v q_m}
\d_{\v 0, \sum \v p_i+\sum \v q_j}\nn  
\Gamma^{(n,m)}(\v p_1,..,\v p_n,\v q_1,..,\v q_m)
\k_{\v p_1}..\k_{\v p_n}\u_{\v q_1}..\u_{\v q_m}\ ,\label{3,28}\eea 
and a completely
analogous equation for $\hat\L$.
Translating the evolution equation of $\L$, (\ref{3,2}), in terms of 
the effective vertices $\hat\Gamma$ of $\hat\L$, we get:
\bea \(\la\p_{\la}+4-n-2m\)
\hat\Gamma^{(n,m)}(\v p_1,..,\v p_n,\v q_1,..,\v q_m)=\nn
\la^2\sum_{\v p}{1\o2}\la\p_{\la}k^2(\({p\o\la}\)^2)C^{-1}(p)
\sum_{k=1}^n\sum_{l=1}^m
\sum_{[i_1,..,i_k][i_{k+1},..,i_n]}\sum_{[j_1,..,j_l][j_{l+1},..,j_m]}
\d_{\v p, \sum \v p_i+\sum \v q_j}\nn  
\hat\Gamma^{(k+1,l)}\(\v p,\v p_{i_1},..,\v p_{i_k},\v q_{j_1},..,\v q_{j_l}\)
\hat\Gamma^{(n-k+1,m-l)}\(-\v p,\v p_{i_{l+1}},..,\v p_{i_n},
\v q_{j_{l+1}},..,\v q_{j_m}\)-\nn
 \sum_{\v p}{1\o2}\la\p_{\la}k^2(\({p\o\la}\)^2)C^{-1}(p){1\o\la^2 L^4}
\hat\Gamma^{(n+2,m)}(\v p,-\v p,\v p_1,..,\v p_n,\v q_1,..,\v q_m)\ 
.\label{3,29}\eea

For $\la>>{1\o L}$ the sums in the 
right-hand side of (\ref{3,29}) can be safely replaced with integrals whose
measure is: $d^4 p\({L\o 2\pi}\)^4$. In the same limit the Kronecker
$\d_{\v p, \sum \v p_i+\sum \v q_j}$ has to be replaced by the Dirac
measure:
$\({2\pi\o L}\)^4\d (\v p- \sum \v p_i-\sum \v q_j)$. 
This yields: 
\bea \(\la\p_{\la}+4-n-2m\)
\hat\Gamma^{(n,m)}(\v p_1,..,\v p_n,\v q_1,..,\v q_m)=\nn
{1\o2}\int {d^4p\o (2\pi\la)^4}\ \v p\cdot\v \p_{p}k^2\(\({p\o\la}\)^2\)
{\la^2\o p^2+m^2}
\hat\Gamma^{(n+2,m)}(\v p,-\v p,\v p_1,..,\v p_n,\v q_1,..,\v q_m)-\nn
{1\o2}\int d^4p\ \v p\cdot\v \p_{p}k^2\(\({p\o\la}\)^2\)
{\la^2\o p^2+m^2}\sum_{k=1}^n\sum_{l=1}^m
\sum_{[i_1,..,i_k][i_{k+1},..,i_n]}\nn
\sum_{[j_1,..,j_l][j_{l+1},..,j_m]}
\d (\v p- \sum \v p_i-\sum \v q_j) 
\hat\Gamma^{(k+1,l)}\(\v p,\v p_{i_1},..,\v p_{i_k},\v q_{j_1},..,\v 
q_{j_l}\)\nn
\hat\Gamma^{(n-k+1,m-l)}\(-\v p,\v p_{i_{l+1}},..,\v p_{i_n},
\v q_{j_{l+1}},..,\v q_{j_m}\)\ .\label{3,30}\eea
Now, rescaling the integration variable, by setting:
\be \hat\Gamma^{(n,m)}(\v p_1,..,\v p_n,\v q_1,..,\v q_m)=
\gamma^{(n,m)}\({\v p_1\o\la},..,{\v p_n\o\la},{\v q_1\o\la},..,{\v q_m\o\la}\)
\ ,\label{3,31}\ee  we can make the right-hand 
side of (\ref{3,30}) $\la$-independent
at high momenta ($p^2>>m^2$).

With the purpose of analysing perturbatively our approach, we reintroduce
into the evolution equation for $\gamma^{(n,m)}$
 the loop counting parameter $\hbar$ 
appearing in (\ref{2,32}). This is done multiplying by a factor ${1\o\hbar}$
both the covariance $C$ and the interaction $\L$: we obtain:
\bea \(\la\p_{\la}+4-n-2m-D\)\gamma^{(n,m)}=-{1\o2}
\int d^4p {1\o p^2+{m^2\o\la^2}}\ \v p\cdot\v \p_{p}k^2(p^2)\nn
\sum_{k=1}^n\sum_{l=1}^m
\sum_{[i_1,..,i_k][i_{k+1},..,i_n]}\sum_{[j_1,..,j_l][j_{l+1},..,j_m]}
\d (\v p- \sum \v p_i-\sum \v q_j)\nn 
\gamma^{(k+1,l)}\(\v p,\v p_{i_1},..,\v p_{i_k},\v q_{j_1},..,\v q_{j_l}\)
\gamma^{(n-k+1,m-l)}\(-\v p,\v p_{i_{l+1}},..,\v p_{i_n},
\v q_{j_{l+1}},..,\v q_{j_m}\)+\nn  
{\hbar\o2}\int {d^4p\o (2\pi)^4}\ \v p\cdot\v \p_{p}k^2(p^2)
{1\o p^2+{m^2\o\la^2}}
\gamma^{(n+2,m)}(\v p,-\v p,\v p_1,..,\v p_n,\v q_1,..,\v q_m)\ 
,\label{3,32}\eea
where $D$ is the dilation operator:
\be D=\sum_i\v p_i\cdot\v \p_{p_i}+
\sum_j\v q_j\cdot\v \p_{q_j}\ .\label{3,33}\ee

To proceed further, it is convenient to introduce the functional generator
$\ll$ of the vertices $\gamma^{(n,m)}$, that is:
\bea \ll [\f,\u]=\sum_{n=0,m=0}^{\infty}{1\o n!}\int\prod_{i=1}^n\(d^4 p_i \f  
(\vec p_i)\)\nn\prod_{j=1}^m\(d^4 q_j \u (\vec q_j)\)
(2\pi)^4\d ( \sum \v p_i+\sum \v q_j) \nn 
\gamma^{(n,m)}\(\v p_{i_1},..,\v p_{i_n},\v 
q_{j_1},..,\v q_{j_m}\)\ .\label{3,34}
\eea 
Notice that the whole set of transformations which has led us from $\L$
to $\ll$ consists in the introduction of dimensionless variables combined
with the infinite volume limit. Indeed setting:
\be \f (x)\equiv\la\hat\f(\la x)\ ,\ \u (x)\equiv\la^2\hat\u (\la x)\ 
,\label{3,35}\ee
we have set:
\be \L(\k,\u,\la,\lao,\rho_0)=
\ll\(K_{\la}\hat\f,\hat\u,\la,\lao,\rho_0\)\ .\label{3,36}\ee
Introducing;
\be \D '(p)= \v p\cdot\v \p_{p}k^2(p^2)
{1\o p^2+{m^2\o\la^2}}\ ,\label{3,37}\ee
and the functional differential operators:
\be {\cal N}=\int d^4p\ \f(\vec p){\d\o\d\f(\vec p)}\ ,\label{3,38}\ee
\be {\cal M}=\int d^4p\ \u(\vec p){\d\o\d\u(\vec p)}\ ,\label{3,39}\ee
\be {\cal D}=\int d^4p\[ \f(\vec p) \v p\cdot\v \p_{p}{\d\o\d\f(\vec p)}+
\u(\vec p) \v p\cdot\v \p_{p} {\d\o\d\u(\vec p)}\]\ ,\label{3,40}\ee
we can write (\ref{3,32}) in the form:
\be \[ \la{\p\o\p\la}-{\cal N}-2{\cal M}-{\cal D}\]e^{-\ll\o\hbar}=
{\hbar\o2}\ipp{\D '(p)\o (2\pi)^4}
{\d\o\d\f(\vec p)}{\d\o\d\f(-\vec p)}e^{-\ll\o\hbar}\ .\label{3,41}\ee
 Representing graphically  by blobs (vertices) with $n$ legs 
the $n$-th functional derivative of $\ll$  and with a solid line 
$\D ' (p)$, we see that the first term in the 
right-hand side of (\ref{3,32}) can be represented by the fusion of two vertices
into a single one joining two lines, while, in the second term, two lines
of the same vertex joining together form a loop.
\be \centerline{
\begin{picture}(250,60)(0,0)
\put(5,30){\makebox(0,0){$\left[ \Lambda{\partial\over\partial\Lambda}
-{\cal N}-2{\cal M}-{\cal D}\right]$}}
\put(76,30){\circle*{12}}
\put(125,30){\makebox(0,0){$=-{1\over2}\int {d^4p\over (2\pi )^4}\left
[\right.$}}
\put(170,22){\line(0,1){16}}\put(170,16){\circle*{12}}\put(170,44){\circle*{12}}
\put(190,30){\makebox(0,0){$-\hbar$}}
\put(220,30){\circle{24}}\put(232,30){\circle*{12}}
\put(244,30){\makebox(0,0){$\left.\right]$}}
\end{picture}}\ .\ee 

The infinite volume limit and momentum scale transformations performed above
can be also applied to $\p_{\rho_a}\L$ and to $V(\la)$ giving $\p_{\rho_a}\ll$
and:\be v(\la)=\lao\p_{\lao}\ll\vert_{\rho, \ \la}\ .\label{3,42}\ee
The evolution equations of these functionals are easily deduced from 
(\ref{3,19})
and (\ref{3,25}), by replacing $\la\p_{\la}$ with 
$\la{\p\o\p\la}-{\cal N}-2{\cal M}-{\cal D}$ and the operator $M$ with
$m$ defined by:
\be m[X]\equiv \ipp{\D '(p)\o (2\pi)^4}\[
{\d\o\d\f(\vec p)}X{\d\o\d\f(-\vec p)}\ll-{\hbar\o2}
{\d\o\d\f(\vec p)}{\d\o\d\f(-\vec p)}X\]\ .\label{3,43}\ee
The normalization functionals $N_a$ are also changed, owing to the fact
that, after the infinite volume limit, the momentum $\v p$ is no longer
an index but a continuous variable. One has also to take into account that
both the fields and the effective lagrangian have been rescaled.
Therefore, after this limit, one has in particular:
\be N_1=2^8\ipp{1\o (2\pi)^4}{\d^4\o\d\f^4(\vec p)}\ ,\label{3,44}\ee
\be N_2=2\v \p_q\cdot\v \p_q\ipp{1\o (2\pi)^4}
{\d\o\d\f(\vec p+\v q)}{\d\o\d\f(\vec p-\v q)}\vert_{\v q=0}\ ,\label{3,45}\ee
while one has:
\be N_3=2^4\la^2\ipp{1\o (2\pi)^4}{\d^2\o\d\f^2(\vec p)}\equiv
\la^2 n_3\ ,\label{3,46}\ee
and in general a factor $\la^{d_a}$ appears in the normalization operator
of the parameter $\rho_a$ with mass dimension $d_a$.

Therefore the normalization conditions become
\be \gamma^{(4,0)}(0) =\rho_1\quad,\quad\p_{p^2}\gamma^{(2,0)}(0) =
\rho_2
\quad,\quad\gamma^{(2,0)}(0) ={\rho_1\o\Lambda^2}\ ,\ee 
and
\be \gamma^{(2,1)}(0) =2\rho_4\quad,\quad\gamma^{(0,2)}(0) =2\rho_5
\quad,\quad\gamma^{(0,1)}(0) ={\rho_6\o\Lambda^2}\ .\ee 
Unluckily, until now, nobody has been able to solve our equations in any
interesting case in 4 dimensions. The analytical Wilson renormalization
group method has been successfully applied only "near" two dimensions 
\cite{7} .
What we are going to do in the following is to apply to our equations 
a purely dimensional analysis and, on this basis, to discuss their pertubative
solution.
\vfill\eject
\salto
\sec{Analysis of the perturbative solution.}
\salto

We shall now briefly discuss the perturbative solution to (\ref{3,41}).

In the perturbative framework the effective lagrangian and its initial 
value $L_0$ are considered as a formal power series in $\hbar$. Therefore in
particular the bare constants $\rho_0$ and the vertices $\Gamma$,
$\hat\Gamma$ and $\gamma$ are formal power series. In this framework the
solution to (\ref{3,41}) can be directly constructed as follows.

We notice, first of all, that the initial value $ l_0(\lao)$
 of $\ll$ is directly obtainable by
applying to (\ref{2,21}) the transformations (\ref{3,27}) and 
(\ref{3,31}) which yield:
\be  l_0(\la)=\int d^4x \[{\lambda_0\o 4!}\hat\f^4+{z_0-1\o 2}(\p\hat\f)^2+
{m_0^2-m^2\o 2\la^2}\hat\f^2
+\zeta_0\hat\u\hat\f^2+\eta_0\hat\u^2+{\xi_0\o\la^2}\hat\u\]\ ,\label{4,1}\ee
where we have used the Fourier transformed variables: 
\be \hat\f(\vec x)=\ipp e^{-i(\vec p\cdot\vec x)}\f(\vec p)\ .\label{4,2}\ee
We also introduce the cut-off propagator:
\be \D (p)={k^2({p\la\o\lao})-k^2(p)\o p^2+{m^2\o\la^2}}\ .\label{4,3}\ee
Now it is sufficient to notice that:
\be \[ \la{\p\o\p\la}-{\cal N}-2{\cal M}-{\cal D}\]l_0(\la)=0\ ,\label{4,4}\ee
and the commutation relation:
\bea \[\[ \la{\p\o\p\la}-{\cal N}-2{\cal M}-{\cal D}\],
\ipp {\D (p)\o (2\pi)^4} {\d\o\d\f(\vec p)}{\d\o\d\f(-\vec p)}\]=
\nn\ipp {\D '(p)\o (2\pi)^4} 
{\d\o\d\f(\vec p)}{\d\o\d\f(-\vec p)}\ ,\label{4,5}\eea
to prove that;
\be \ll=-\hbar\ ln\[ e^{{\hbar\o2}\ipp {\D (p)\o (2\pi)^4}
{\d\o\d\f(\vec p)}{\d\o\d\f(-\vec p)}}
e^{-{l_0(\la)\o\hbar}}\]\ ,\label{4,6}\ee
is the solution of (\ref{3,41}) with initial value $l_0(\lao)$.
It is also easy to verify that $\ll$ is the functional generator of the
connected and amputated Feynman amplitudes corresponding to the propagator
$\D (p)$ and to the vertices described by $l_0$. 

Using this result one could try to study the existence of an UV limit of the 
perturbative theory. However we shall see in the following that, in order
to have a regular UV limit, we have to consider as fixed
parameters the renormalized $\rho$'s, given in (\ref{3,10}) computed at 
$\la=\la_R$. 

From (\ref{4,6}) we find in particular:
\bea \gamma^{(2,0)}(p^2) ={\rho_{0,2}p^2+{\rho_{0,3}\o\Lambda^2}\o
1+\(\rho_{0,2}p^2+{\rho_{0,3}\o\Lambda^2}\)\D(p^2)}+O(\hbar)\ ,\nn 
\gamma^{(4,0)} =\rho_{0,1}+O(\hbar)\ ,\label{4,6b}\eea 
and
\be \gamma^{(6,0)}\(\vec p_1,..,\vec p_6\)\sum_{i_1\geq i_2 \geq i_3=1}^6
\rho_{0,1}^2
\D\(\(\v p_{i_1}+\v p_{i_2}+\v p_{i_3}\)^2\)\ .\label{4,6c}\ee 
The analysis of the UV limit 
will develop along the following lines.
We shall introduce a suitable class of norms for $\ll$ and 
its $\lao$-derivative $v(\la)$ in the new parametrization. 
We then show that,
for $\lao$ large with respect to $\la_R$ and $\la$ and 
at any finite perturbative order, this functional
vanishes in the UV limit faster than $\({\la\o\lao}\)^{2-\epsilon}$ times $\ll$
for any positive $\epsilon$.
This will immediately imply the UV convergence of 
the theory.
Now (\ref{4,6}) is explicitly parametrized in terms of the
bare constants which, written as functions of $\rho_R$, coincide with 
the counter-terms associated with the corresponding
vertices. These are $\hbar$-ordered series. 
Thus the reparametrization of $\ll$ in terms 
of the renormalized constants
is equivalent to the introduction of a subtraction procedure.   
This is however extremely cumbersome. In particular checking the 
mechanism of subtraction of the UV divergent contributions of
subdiagrams with the corresponding counter-terms, requires a detailed
analysis of the so-called overlapping divergences.
The advantage of the method presented here, which is inspired by the work
of Polchinski, is that, using explicitly the evolution equations 
(\ref{3,2})
and (\ref{3,19}), one can reach the same results without any diagrammatic 
analysis.

Concerning the just mentioned change of parametrization, let us notice 
that, as shown above in the tree approximation, the two-field vertex 
$\gamma^{(2,0)}$ involves tree parameters, $m$, $\rho_2$ and $\rho_3$. 
However the quantity that is physically relevant  is the correlation 
length that in our approach turns out to be a function of  the above 
mentioned parameters. This seems to leave a wide 
freedom in particular in the choice of $\rho_2$ and $\rho_3$ that 
can be compensated  by that of the covariance $m^2+p^2$, leaving the 
correlation length unchanged. 

However we shall see in a moment that in a loop ordered perturbation theory, 
to simplify the recursive solution of the evolution 
equations, it is highly convenient 
to choose the two-field vertex $\g^{(2,0)}$ vanishing 
at the tree level.  This means that $\rho_2$ and 
$\rho_3$ are not "really" free parameters. They must be of order $\hbar$.
This assumption is perfectly compatible with our evolution equations from 
which we can compute, in the case of a sharp cut-off
\be \rho_{2}=\rho_{0,2}-{\hbar\o 2(4\pi)^2}\rho_{0,1}\ ,\label{4,6d}\ee 
\be \rho_{3}=\rho_{0,3}+O(\hbar^2)\ ,\label{4,6e}\ee 
\be \rho_{1}=\rho_{0,1}+O(\hbar)\rho_{0,1}^2\ .\label{4,6f}\ee 
It is also clear from (\ref{4,6b}) and this result that the jacobian matrix 
$\({\p \rho\o\p\rho_0}\)$ and the functional $\p_{\rho}\ll$ remain 
perfectly regular after this choice.

Owing to the nature of the normalization conditions, we have 
to study the $\la$ dependence of  the vertices generated by $\ll$, 
$\p_{\rho_a}\ll$ and $v$ together with their momentum derivatives.

 The evolution equations for the momentum derivatives of the vertices are
immediately deducible from (\ref{3,32}). One can verify in particular from 
this
equation how the derivatives distribute in its the right-hand side. 
For purely formal reasons we briefly discuss here how the insertion of
momentum derivatives can be performed at the functional level. This point, 
that
is not crucial to the study of the perturbative UV limit, is motivated by 
the
formal need to translate the whole normalization group action into a finite
number of differential equations for the functional generators.

To define these momentum derivatives within the functional framework
we introduce the family of differential operators:
\be D_{\nu,n,m}\(\v p_1,...,\v p_{n+m}\)=
P_{\nu}\(\v \p_{ p_k}-\v \p_{p_l}\)
\prod_{i=1}^n{\d\o\d\f (\v p_i)}
\prod_{j=n+1}^{n+m}{\d\o\d\u (\v p_j)}\ ,\label{4,7}\ee
where $P_{\nu}$ is a polynomial of degree $d(\nu)$. 
Considering in particular the evolution equations for $D_{\nu} \ll$ one 
finds that, due to the non-linearity of (\ref{3,32}), these equations contain a
great number of terms. Let us consider for example:
\be \v D_{1,2,0}\(\v p_1,\v p_2\)=\(\v \p_{p_1}-\v \p_{p_2}\){\d\o\d\f (\v p_1)}
{\d\o\d\f (\v p_2)}\ .\label{4,8}\ee
and notice that for any pair of 
functionals ${\cal A}$ and ${\cal B}$ one has:
\bea \v D_{1,2,0}\(\v p_1,\v p_2\)\ippp{\df\v p)}{\cal A}{\df-\v p)}{\cal 
B}=\nn 
\ippp{\df\v p)}\v D_{1,2,0}\(\v p_1,\v p_2\){\cal A}{\df-\v p)}{\cal B}+\nn
\ippp\v D_{1,2,0}\(\v p_1,\v p\){\cal A}{\df\v p_2)}{\df-\v p)}{\cal B}
+\nn 
\ippp{\df\v p_1)}{\df\v p)}{\cal A}\v D_{1,2,0}\(-\v p,\v p_2\){\cal B}-\nn
\ipp \v\p_p {\D '(p)\o (2\pi)^4}{\df\v p_1)}{\df\v p)}{\cal A}
{\df\v p_2)}{\df-\v p)}{\cal B}+\({\cal B}\leftrightarrow {\cal A}\)\ 
.\label{4,9}\eea
Iterating the same equation, we see that for a generic operator $D_{\nu}$:
\bea D_{\nu,n,m}\ippp{\df\v p)}{\cal A}{\df-\v p)}{\cal B}=\nn  
\sum_{\rho,\sigma,\tau,k,l}\ipp P_{\rho}(\v\p_p)
 {\D '(p)\o (2\pi)^4}D_{\sigma,k+1,l}{\cal A}D_{\tau,n+1-k,m-l}{\cal B}\ ,
\label{4,10}\eea
where the sum is restricted to the differential operators and polynomials
for which: \be d(\rho)+d(\sigma)+d(\tau)=d(\nu)\ .\label{4,11}\ee

We have thus the evolution equation:
\bea \[ \la{\p\o\p\la}-{\cal N}-2{\cal M}-{\cal D}-n-2m-d(\nu)\]D_{\nu,n,m}\ll
=\nn  {\hbar\o2}\ippp D_{\nu,n,m}{\df\v p)}{\df-\v p)}\ll-\nn
{1\o2}\sum_{\rho,\sigma,\tau,k,l}\ipp P_{\rho}(\v\p_p)
 {\D '(p)\o (2\pi)^4}D_{\sigma,k+1,l}\ll D_{\tau,n+1-k,m-l}\ll\ 
,\label{4,12}\eea
where again the sum in the right-hand side is restricted by (\ref{4,11}).
Analogous equations hold true for $\p_{\rho_a}\ll$ and $v$.

This completes the construction of the evolution equations. To study their
solutions we have to translate these 
evolution equations for the generating functional into
those for the corresponding $N$ field and $M$ operator vertices. 
For this we have to clarify a technical point. Since the functional 
generators
are translation invariant, the corresponding vertices are proportional to a
Dirac delta function in the sum of the external leg momenta. To get rid of 
this delta function we notice
that, given any translation invariant functional $\cal A$, the
$(\nu,n,m)$ momentum derivative of the corresponding $(N,M)$ vertex is given 
by
\bea D_{\nu,n,m}{\cal A}^{(N,M)}\(\v p_1,...,\v p_{n+m}\)=
(N+M)^4\int {d^4k\o(2\pi)^4}\prod_{i=n+m}^{n+m+N}{\d\o\d\f (\v p_i+\v 
k)}\nn
\prod_{j=n+m+N+1}^{n+m+N+M}{\d\o\d\u (\v p_j+\v k)}
D_{\nu,n,m}\(\v p_1+\v k,...,\v p_{n+m}+\v k\){\cal A}\vert_{\f=\u=0}\ .
\label{4,13}\eea 
For these vertices we define the norms:
\be \Vert D_{\nu,n,m}{\cal A}\Vert^{(N,M)}\equiv sup_{p_i^2<2, \sum \v p_i=0}
\vert D_{\nu,n,m}{\cal A}^{(N,M)}\vert\ .\label{4,14}\ee
The value of these norms in the tree approximation can be easily computed
from (\ref{4,6}), (\ref{4,6b}) and (\ref{4,6c}) from which it turns
out that the norms of $\ll$ can be bounded 
above and below by constants for a generic choice of the parameters $\rho$ 
and for:
\be \la_R\leq\la<<\lao\ .\label{4,15}\ee Furthermore from (\ref{4,6}) and
(3,42) one easily finds that:
\be \Vert v(\la)\Vert^{(N,M)}<
C_{N,M} \({\la\o\lao}\)^2\Vert\ll\Vert^{(N,M)}\ .\label{4,17}\ee
Notice that the condition $\lao>>\la$ is needed to 
guarantee the assumed lower bound on the 
vertices; indeed when $\la$ tends to $\lao$ almost all of them vanish
together with the propagator (\ref{4,3}). In the following we shall 
disregard (\ref{4,15}) whenever we discuss upper bounds.

To push our analysis to all orders, the first step is to evaluate
the norms of $\ll$ and $\p_{\rho}\ll$. We shall then discuss those of $v$.

We study the evolution equations recursively for 
increasing order ($k$) in $\hbar$ and degrees (N and M) in the fields.

From now on we shall label by the index $k$ the $k^{th}$ order term of a 
vertex.

For $\ll$ we assume the induction
hypothesis that:
\be \Vert D_{\nu}\c_k\Vert^{(N,M)}<P_k\(ln\({\la\o\la_R}\)\)\ ,\label{4,18}\ee
where $P_k$ is a polynomial of degree $k$ with positive cofficients.
 (\ref{4,18}) is, of course, verified for $k=0$.

Let us notice now that the essential 
simplification introduced by perturbation theory
consists in the possibility of performing a recursive analysis of 
(\ref{3,32})
for increasing loop order and for any given order in $\hbar$ increasing
$N$ and $M$ starting from $(2,0)$. This is possible since
the right-hand side of the evolution equation for the $(N,M)$ vertex
at the perturbative order $k$ depends 
 on the $(N-2,M)$ and $(N,M-1)$ vertices at the same
order, while the other vertices appearing in it have lower orders. Indeed,
as already noticed, the first term in the right-hand side of (\ref{3,32}) 
corresponds to the product of two vertices joined by one line, this has a
tree graph structure, hence the sum of the orders of the two vertices is $k$,
the total number of $\f$-legs is $N+2$, that of $\u$-legs is $M$. 
Owing to the particular choice of the parametrization discussed above, which
implies the vanishing of the $(2,0)$ vertex in the tree approximation, this 
first term does not contain vertices of order $k$ with more than $N-1$
$\f$-legs and $M$ $\u$-legs. In the second term in the right-hand side of 
(3,32) there appears only the $(N+2,M)$ vertex at the order $k-1$.

A further remark which is essential to obtain an upper bound for the 
absolute value of the right-hand side of (\ref{3,32}) at each recursive 
step, is that $\D '(p)$ is absolutely bounded together
with all its derivatives and that it vanishes identically 
outside the hypersphere of radius 2 and inside that of radius 1.  

Taking into account these remarks and (\ref{4,18}) we conclude that at the
order $k$ the absolute value of the right-hand side of (\ref{3,32}) is bounded
by a polynomial of degree $k$ in $ln\({\la\o\la_R}\)$.

Coming to the study of the single vertices, let us consider, first of all:
\bea D_4^{\mu\nu\rho\sigma}\c_{k+1}^{(2,0)}\equiv
\p_p^{\mu}\p_p^{\nu}\p_p^{\rho}\p_p^{\sigma}\c_{k+1}^{(2,0)}(\v p)=\nn
4\(\d^{\mu\nu}\d^{\rho\sigma}+\d^{\mu\rho}\d^{\sigma\nu}+
\d^{\mu\sigma}\d^{\rho\nu}\)\c_{k+1}^{(2,0)}(p^2)^{(ii)}+\nn  
8\(\d^{\mu\nu}p^{\rho}p^{\sigma}+\d^{\mu\rho}p^{\sigma}p^{\nu}+
\d^{\mu\sigma}p^{\rho}p^{\nu}+\d^{\rho\sigma}p^{\mu}p^{\nu}
+\d^{\sigma\nu}p^{\mu}p^{\rho}+\right.\nn\left.\d^{\rho\nu}p^{\mu}p^{\sigma}\)
\c_{k+1}^{(2,0)}(p^2)^{(iii)}+
16p^{\mu}p^{\nu}p^{\rho}p^{\sigma}\c_{k+1}^{(2,0)}(p^2)^{(iv)}\ 
,\label{4,19}\eea
where $\c_{k+1}^{(2,0)}(p^2)^{(x)}$ stays for the $x^{th}$ derivative
of $\c_{k+1}^{(2,0)}(p^2)$ with respect to $p^2$.
We have:
\be \Vert \(\la{\p\o\p\la}- D-2\)D_4\c_{k+1}\Vert^{(2,0)}<
P_{k+1}\(ln\({\la\o\la_R}\)\)\ ,\label{4,20}\ee which has to be integrated with
the initial condition: 
\be D_4\c_{k+1}^{(2,0)}\vert_{\la=\lao}=0\ .\label{4,21}\ee
Now we can integrate (\ref{4,21}) as follows. Setting:
\be D_4\c_{k+1}^{(2,0)}(\v p)=\({\la\o\lao}\)^2F(\la\v p,\la)\ ,\label{4,22}\ee
one has:
\be \vert\la{\p\o\p\la}F\vert <\({\lao\o\la}\)^2P_{k+1}\(ln\({\la\o\la_R}\)\)
\ ,\label{4,23}\ee
then, integrating, one gets:
\be \vert F\vert <\({\lao\o\la}\)^2P'_{k+1}\(ln\({\la\o\la_R}\)\)\ 
,\label{4,24}\ee
and hence one recovers (\ref{4,18}) for $D_4\c_{k+1}^{(2,0)}$.

It is important to notice, at this point, that the initial condition 
(\ref{4,21})
has not been relevant for the final result (\ref{4,18}). This is 
due to the fact that
the initial condition contributes to the solution of (\ref{4,12}) proportionally
to that of the associated homogeneous equation (the one with vanishing
right-hand side). Considering a momentum derivative of order $d$ of the vertex
$(N,M)$ this homogeneous solution is proportional to $\({\la\o\lao}\)^
{N+2M+d-4}$: it is therefore irrelevant if the exponent is positive.
Now we compute:
\bea {1\o3!}\int dt (1-t)^3p_{\mu}p_{\nu}p_{\rho}p_{\sigma}
D_4^{\mu\nu\rho\sigma}\c_{k+1}^{(2,0)}(t\v p)=\nn  \c_{k+1}^{(2,0)}(p^2)-
\c_{k+1}^{(2,0)}(0)-p^2\c_{k+1}^{(2,0)(i)}(0)\ .\label{4,25}\eea
Notice that, independently of the presence of a mass, our vertices 
are not affected with infra-red singularities since the 
propagator (\ref{4,3}) is cut-off both at high and low momenta.

Then we get:
\be \Vert \c_{k+1}\Vert ^{(2,0)}< P_{k}'\(ln\({\lao\o\la}\)\)
+\vert \c_{k+1}^{(2,0)}(0)\vert
+4\vert \c_{k+1}^{(2,0)(i)}(0)\vert\ .\label{4,26}\ee
The second and third term in the right-hand side can be computed
using again the evolution equations. However now these equations have to 
be integrated between $\la_R$ and $\la$ since the value of both terms
is fixed by the renormalization conditions (\ref{3,10}) at $\la_R$.

In much the same way as above we consider:
\be D_2^{\mu\nu}\c_{k+1}^{(2,0)}(0)\equiv \p_p^{\mu}\p_p^{\nu}
\c_{k+1}^{(2,0)}(0)=2\d^{\mu\nu}\c_{k+1}^{(2,0)(i)}(0)\ ,\label{4,27}\ee
for which the evolution equation gives:
\be \vert \la{\p\o\p\la}D_2\c_{k+1}^{(2,0)}(0)\vert<
P_{k}\(ln\({\la\o\la_R}\)\)\ ,\label{4,28}\ee  T
he right-hand side of this equation receives contributions only
from the term proportional to $\hbar$ in (\ref{3,32}) since $\D '(p)$
vanishes identically for $p<1$. 

Integrating (\ref{4,28}) between 
$\la_R$ and $\la$ and taking into account (\ref{4,27}) yields:
\be \vert\c_{k+1}^{(2,0)(i)}(0)\vert<P_{k+1}\(ln\({\la\o\la_R}\)\)\ 
.\label{4,29}\ee
Notice that the degree of the polynomial $P$ has increased by one.
This will happen every time we shall integrate a vertex evolution equation
whose left-hand side is of the form of that in (\ref{4,28}).

In an analogous way we have:
\be \vert \(\la{\p\o\p\la}+2\)\c_{k+1}^{(2,0)}(0)\vert<
P_{k}\(ln\({\la\o\la_R}\)\)\ ,\label{4,30}\ee leading, after integration, to:
\be \vert\c_{k+1}^{(2,0)}(0)\vert<P"_{k}\(ln\({\la\o\la_R}\)\)\ 
.\label{4,31}\ee
Notice that here the normalization at $\la_R$ has an irrelevant influence
on the value of the right-hand side.

Now, combining (\ref{4,26}), 
(\ref{4,29}) and (\ref{4,31}), we get back the induction 
hypothesis at order $k+1$ for $\Vert \ll\Vert ^{(2,0)}$.

We have discussed in some detail this step of the recursive proof
to give an example of the general strategy, which is essentially the
following. For every (N,M) we consider a momentum-derivative of the vertex
of  high enough order
to make it "convergent" (N+2M+the order of derivation larger than 4)
and we compute this momentum derivative integrating the evolution equation
between $\la$ and $\lao$ where it has to vanish.

The original vertex, or its derivatives of low order, are then computed
from the derivative of higher order integrating it with respect to
a momentum scale variable, as in (\ref{4,25}).
The resulting formula contains integration
constants, corresponding to the values at zero momentum of the vertex and
its derivatives of low order (its "divergent part"). 
These are computed integrating their evolution
equations between $\la$ and $\la_R$ where they are assigned by the 
normalization conditions.

The procedure turns out to be much simpler in the case of "convergent"
vertices $(N+2M>4)$ which are determined directly by integrating the evolution
equation between $\la$ and $\lao$ and in the case $(N=0\ ,\ M=1)$ where it is 
sufficient to integrate between $\la$ and $\la_R$.

The results that we have so far reached concerning $\ll$ could easily be 
resumed
formulating a theorem. We avoid this assuming that the crucial steps of our
study have been sufficiently clarified.

The above method can be employed to evaluate the norms of $\p_{\rho}\ll$
for which we assume the induction hypothesis that:
\be \Vert\p_{\rho_a} D_{\nu}\c_k\Vert^{(N,M)}<
\la^{-d_a}P_{k}\(ln\({\la\o\la_R}\)\)\ ,\label{4,32}\ee
where, as mentioned above, $d_a$ is the mass dimension of the 
parameter $\rho_a$.

The validity of (\ref{4,32}) in the tree approximation can be read directly from
(4,1). To extend (\ref{4,32}) 
to all orders we repeat the analysis described above for
$\ll$ noticing that the initial conditions at $\la_R$ and $\lao$ for
$\p_{\rho_a}\ll$ can be easily deduced from those for $\ll$.

After infinite volume limit and momentum rescaling 
 the evolution equation for $\p_{\rho}\ll$ becomes:
\bea \[ \la{\p\o\p\la}-{\cal N}-2{\cal M}-{\cal D}\]\p_{\rho_a}\ll=\nn  
m[\p_{\rho_a}\ll] -\la^{-d_b}n_b\[m[\p_{\rho_a}\ll]\]\p_{\rho_b}\ll\ .\label{
4,33}\eea
Let us remark that in the second term of the right-hand side of this equation
the factor $\la^{-d_b}$ coming from the normalization operator compensates
$\la^{d_b}$ appearing in (\ref{4,32}). Therefore the whole right-hand side
turns out to be proportional to $\la^{-d_a}$.

As above we start from the study of $\p_{\rho_a}D_4\ll$ for which we have:
\be \Vert \(\la{\p\o\p\la}- D-2\)\p_{\rho_a}D_4\ll\Vert^{(2,0)}<\la^{-d_a}
P_{k+1}\(ln\({\la\o\la_R}\)\)\ .\label{4,34}\ee The integration between 
$\la$ and $\lao$ (\ref{4,34}) leads back to (\ref{4,32}) 
for $\p_{\rho_a}D_4\ll$. 
It is now evident that if
$\rho_a$ is a dimensionless parameter ($d_a=0$) the recursive proof of 
(\ref{4,32})
is identical to that of (\ref{4,18}). However, if $d_a>0\ ,$ we have to change 
the strategy of our proof. Indeed, for example, the evolution 
equation of $\p_{\rho_a}D_2\c_{k+1}^{(2,0)}(0)$ yields:
\be \vert \la{\p\o\p\la}\p_{\rho_a}D_2\c_{k+1}^{(2,0)}(0)\vert<\la^{-d_a}
P_{k+1}\(ln\({\la\o\la_R}\)\)\ .\label{4,35}\ee 
Comparing this with (\ref{4,22}) and (\ref{4,28}) we see that integrating 
down from $\lao$ we can profit of the irrelevancy of the high energy value 
in the present situation. Thus we get:
\be \vert\p_{\rho_a}D_2\c_{k+1}^{(2,0)}(0)\vert< \la^{-d_a} 
P_{k+1}'\(ln\({\la\o\la_R}\)\)\ .\label{4,36}\ee 
In much the same way we can show that:
\be \vert\p_{\rho_a}\c_{k+1}^{(2,0)}(0)\vert< \la^{-d_a}
P_{k+1}"\(ln\({\la\o\la_R}\)\)\ .\label{4,37}\ee 
Finally, using all these results
to compute $\p_{\rho_a}\c_{k+1}^{(2,0)}(p^2)$, by means of an integration
 procedure completely  analogous to
(4,25), we prove the validity of the induction hypothesis for this derivative.
Following the same strategy, one can study $\p_{\rho_a}\c_{k+1}^{(4,0)}$
starting from a suitable second derivative. 

Then, increasing step by step $N$ and $M$, one can prove (\ref{4,32}) for all 
the coefficients of $\p_{\rho_a}\ll$ at the order $k+1$ completing the 
proof of this hypothesis.

We now come to the study of $v$ defined in (\ref{3,42}) 
and notice, first of all,
that in its evolution equation every factor $ \la^{d_a}$ coming from
a normalization operator compensates the one appearing in the norms
of $\p_{\rho_a}\ll$.

In much the same way
as in the already discussed cases, the analysis begins in the tree 
approximation where $v$ can be computed directly from (\ref{3,42}) and 
(\ref{4,6})-(\ref{4,6f}).
Indeed, in this approximation, $\ll$ depends on $\lao$ only through
the propagator (\ref{4,3}) since the bare and renormalized parameters 
coincide.
Now we have:\be \lao{\p\o\p\lao}\D (p^2)=-{1\o p^2+{m^2\o\la^2}}
\v p\cdot\v \p_{p}k^2\(\({p\la\o\lao}\)^2\)=
-\({\la\o\lao}\)^2\D '\({p\la\o\lao}\)\ ,\label{4,38}\ee thus:
\be sup\vert \lao{\p\o\p\lao}\D\vert=\({\la\o\lao}\)^2 sup \vert\D '\vert
\ .\label{4,39}\ee
One has also to notice that a cut at low momenta appears in 
$\lao{\p\o\p\lao}\D$ at $p={\lao\o\la}$ (while in $\D$ it appears at $p=1$),
therefore the $\lao$-derivative of a vertex vanishes unless the momenta of 
a great number of external momenta add together to overcome this cut. Thus this
derivative does not vanish only for vertices whose external legs increase
proportionally to ${\lao\o\la}$.

Anyway, recalling the structure of the tree diagrams and (\ref{4,34}), we can 
conclude that:
\be \Vert D_{\nu}v_0\Vert^{(N,M)}<C_{N,M,\nu} 
\({\la\o\lao}\)^2\ .\label{4,40}\ee
In order to study the higher perturbative orders using, as above, the 
evolution equation, we have to discuss, first of all, the initial condition
at $\la=\lao$. Before the infinite volume limit this is given in 
(\ref{3,22}).
After this limit and the rescaling of momenta shown in (\ref{3,31}) we have:
\bea v(\lao)={1\o2}\ipp{\D '(p)\o (2\pi)^4}\[
{\d\o\d\f(\vec p)}l_0(\rho_0){\d\o\d\f(-\vec p)}l_0(\rho_0)-\right.\nn
\left. N_a\[{\d\o\d\f(\vec p)}l_0(\rho_0){\d\o\d\f(-\vec p)}l_0(\rho_0)\]
\p_{\rho_a}l_0(\rho_0)\]+\hbar C\ ,\label{4,41}\eea where 
$C$ is field independent.
Recalling (\ref{4,18}) we find for this initial condition the upper bound:
\be \Vert D_{\nu}v_k(\lao)\Vert^{(N,M)}< 
P_k\(ln\({\lao\o\la_R}\)\)\ .\label{4,42}\ee
It is now possible to apply the evolution equations for $v$ repeating the 
analysis already developed for $\ll$ and $\p_{\rho_a}\ll$. In the present
case the recursive procedure will be based on the induction hypothesis:
\be \Vert D_{\nu}v_k\Vert^{(N,M)}<\({\la\o\lao}\)^2
P_k\(ln\({\lao\o\la_R}\)\)\ ,\label{4,43}\ee which is suggested by 
(\ref{4,40}) and 
by the fact that the initial condition at $\lao$ gives irrelevant contributions
to $v$ (in this case irrelevant means of order $\({\la\o\lao}\)^2$).

Considering the right-hand side of the evolution equation 
of $v$, that has essentially the same structure of that of $\ll$ but is 
linear in $v$, we find that:
\bea \Vert \[ \la{\p\o\p\la}-{\cal N}-2{\cal M}-{\cal D}
-n-2m-d(\nu)\]D_{\nu,n,m}
v_{k+1}\Vert^{(N,M)}<\nn  \({\la\o\lao}\)^2
P_{k+1}\(ln\({\lao\o\la_R}\)\)\ .\label{4,44}\eea 
We then begin, order by order and in complete analogy with $\ll$, from
$D_4v_k$, with $D_4$ defined in (\ref{4,19}), and we find:
\be \Vert D_4
v_{k+1}\Vert^{(N,M)}<\({\la\o\lao}\)^2
P_{k+1}'\(ln\({\lao\o\la_R}\)\)\ .\label{4,45}\ee Notice that here the 
initial value at $\lao$ gives a relevant contribution to (\ref{4,45}).

Applying on $D_4v_{k+1}^{(2,0)}$ the integration 
procedure exhibited in (\ref{4,25}) we find
directly $\Vert v_{k+1}\Vert^{(2,0)}$ since $v_{k+1}^{(2,0)}(0)$ and
$v_{k+1}^{(2,0)(i)}(0)$ vanish. Indeed from 
(\ref{3,12})
one finds
\be N_a V(\la)=0\ .\label{4,46}\ee

Then, step by step, we consider the coefficients of $v$ with increasing $N$
and $M$ completing the proof of (\ref{4,43}).

From this inequality we can prove the existence of an UV limit for the 
theory. Indeed from the definition of $v$ we have:
\bea \ll\(\la_R,\lao,\rho_0\(\la_R,\lao,\rho_R\)\)-
\ll\(\la_R,\lao\prime,\rho_0\(\la_R,\lao\prime,\rho_R\)\)=\nn  
\int_{\lao}^{\lao\prime} {dx\o x} v\(\la_R,x,\rho_0\(\la_R,x,\rho_R\)\)\ 
.\label{4,47}\eea
This equation can be translated in terms of the coefficients and, using 
(4,43),
it leads to:
\bea \Vert\ll\[\la_R,\lao,\rho_0(\la_R,\lao,\rho_R)\]
-\ll\[\la_R,\lao\prime,\rho_0(\la_R,\lao\prime,\rho_R)\]\Vert^{(N,M)}<\nn
\vert\lao^2-\lao\prime^2\vert {\la_R^2\o\lao^4}
P_r\(ln\({\lao\o\la_R}\)\)\ ,\label{4,48}\eea
which proves, to any perturbative order $r$, the wanted convergence 
property. Indeed (\ref{4,48}) shows that, if the effective parameters 
$\rho_{R,a}$ are kept fixed, the vertices of the effective theory satisfy
the Cauchy convergence criterion for $\lao\to\infty$.

A further comment about the nature of the UV limit is here possible
if we translate (\ref{4,18}) in terms of the effective vertices $\g$ in
(\ref{3,28}).
Indeed we have that, for $p<\la$ and $q<\la$ the coefficients in the
expansion  of $\L$ in series of $\k$ and $\u$ satisfy, at the order r: 
\be \vert\Gamma_r^{(n,m)}\vert\leq P_r\(log\({\la\o\la_R}\)\) 
(\la)^{4-n-2m}\ .\label{4,49}\ee
For the momentum derivatives of order $\nu$ of these vertices the power 
behavior in the right-hand side of (\ref{4,49}) is increased by $\nu$.

From this inequality we can see in particular the effect of the introduction
into $L_0$ of terms with dimension $d$ higher that 4 (e. g. 6). 
Indeed these terms can be introduced first as composite  operators together
with their sources of negative dimension (e. g. -2) and then they 
can be inserted into the bare interaction by replacing their source with
a coupling coefficient proportional to $\lao^{4-d}$.
Now (\ref{4,49}) shows 
that at the first order in the coupling the effect on $\Gamma^{(n,m)}$ of the 
insertion into the bare lagrangian of an operator
of dimension $d>4$ is proportional to $\la^{d-n-2m}$,
thus, the ratio of this versus the unperturbed effective coefficient vanishes
at low energy as $\({\la\o\lao}\)^{d-4}$. Therefore the effect of a 
"non-renormalizable" term in $L_0$ is "irrelevant" at low energies.

This result agrees with the previous remark that 
in the evolution of "finite" vertices or derivatives of vertices a change in
the initial conditions at $\lao$ gives irrelevant contributions to $\ll$ 
for $\la<<\lao$.

As a further example of irrelevant operators that will be important 
in the study 
of quantum breaking of symmetries. Let us consider a bare operator that in 
the limit $\lao\to\infty$ is local and has finite dimension $d$, while for 
finite $\lao$ it is non-local and contains contributions of any dimension.
Suppose also that the operator satisfies vanishing normalization conditions 
at $\la_R$ at all perturbative orders.
It is possible to show that the operator vanishes in the UV limit.

Indeed  let us consider the evolution equation of the corresponding 
effective operator, the derivative of the effective lagrangian with respect 
to the corresponding source in the origin. The evolution equation is 
directly obtained from (\ref{3,41}) and it is strictly analogous to it. 
However it is linear in the effective operator. This equation can thus be 
recursively solved increasing the dimension and the perturbative order of 
the vertices. In the first step, increasing the dimension of the vertices 
and keeping their order fixed, one begins integrating the evolution equation 
from 
$\la_R$. Due to the vanishing normalization conditions and to the linearity 
of the equation one finds vanishing vertices. However, if the dimension of 
the vertex overcomes that of the operator, one has to integrate down from 
$\lao$ and hence one finds non-vanishing results due to the non-locality of 
the of the bare operator. The vertices of high dimension are however 
proportional to a positive integer power ($n$) of ${\la\o\lao}$ as are the 
irrelevant contributions to the solution of (\ref{3,41}). Thus one finds a 
global upper bound proportional to $\({\la\o\lao}\)^n$ for the vertices of 
the effective operator. It is clear that this upper bound holds true to all 
orders. This proves our claim.

\vfill\eject
\salto
\sec{Composite operators and Wilson expansion.}
\salto

Having so concluded the general study of the UV limit, we have now to deepen
the analysis of composite operators. Until now the 
general method to introduce these operators 
in the framework of the functional technique
has been illustrated referring to the operator $\f^2$ .
The method is based on the introduction of a source
($\u$) of given dimension ($2$). We have also 
seen that, in order to identify completely
an operator one has to assign a suitable set of normalization conditions.
Among these conditions those corresponding to lagrangian terms non-linear in
the source concern possible multi-local contributions to the Green functions
of many composite operators.

We shall now make some comments on this procedure, limiting 
however our attention to the Green functions of a single composite operator
and an arbitrary number of fields.

The first point to be considered is the choice of the dimension of the source.
In our example we have requested the bare lagrangian to have dimension 4 and
therefore we have assigned dimension $2$ 
to the source of an operator of dimension
$2$. Following our study of the effective lagrangian and of the corresponding
evolution equation, it is possible to have a better understanding of this 
choice. Indeed the dimension of the source has influenced its 
rescaling in the infinite momentum limit (\ref{3,31}) and hence 
the evolution equation of $\ll$. On the other hand, in the tree approximation,
the bare structure of the composite operator determines directly the scaling
behavior of the vertices containing its source; this behavior has to match
with the evolution equation in order to be reproduced to all orders of
perturbation theory. In this way we have an upper bound for 
the source dimension.

As we have seen in the analysis of the evolution equation of $\ll$,
once the dimension $\d$ of the source has been fixed, in order to define the
operator completely, we have to assign a normalization condition for every
"divergent" vertex involving the source, that is, for every monomial
in the quantized field and its derivatives with dimension up to $4-\d$,
that of the operator we are willing to define.

Concerning the bare structure of an operator, that fixes 
the initial conditions at $\lao$, it has been
repeatedly remarked that, after infinite volume limit and momentum rescaling,
the addition of terms of higher dimension with $\lao$-dependent coefficients
gives irrelevant contributions to the operator.

Thus, in conclusion, in the UV limit a composite operator is specified by 
assigning its dimension
$4-\d$, or, equivalently, the dimension of its source $\d$,
 and a polynomial operator $P$ with dimension up to $4-\d$ which defines
its normalization conditions at $\la_R$. An 
analogous characterization of composite 
operators has been introduced by Zimmermann \cite{3} who has
represented them with the symbol $N_{4-\d}\[P\]$. 
We can use here the same symbol
provided we take into account 
the normalization point $\la_R$ which in Zimmermann's
definition is equal to zero. 

It is interesting to notice here that, given an operator of dimension $4-\d$,
if one considers only the vertices linear in tis source and if 
multiplies all these vertices by $\({\la\o\la_R}\)^{\eta}$ for $\eta >0$, 
one gets the vertices of a new operator with
dimension $4-\d+\eta$. Indeed the new vertices satisfy the evolution 
equations of the vertices of an operator of this dimension since these 
evolution equations are linear in the operator vertices.
The normalization conditions for the vertices of this new operator can be 
directly computed from the vertices of the original one at $\la_R$. The 
number of the conditions for the new operator is larger than that of the 
original one since the dimension is higher.
Comparing the vertices of the two operators one gets 
a linear relation analogous to that connecting in Zimmermann's scheme
operators with different dimension.

A further important remark concerning our approach to 
composite operators is that nowhere in our analysis the locality of 
the bare operator has played an essential role. Indeed, to identify the 
relevant normalization parameters of a certain operator, we have 
used the evolution equations of its vertices. These equations are 
obtained, independently of the locality of the operator, by taking the 
source derivative of both members of (\ref{3,2}) and hence they 
have the same structure of (\ref{3,14}). Once the evolution equations 
have been determined one has to identify a recursive hypothesis for 
the upper bounds of the vertices. As we have said above these 
bounds are set together with the dimension of the source in the tree 
approximation. Hence, in particular, if a bare operator is non-local
involving fields in  split points at a distance of order ${1\o\lao}$,
in our analysis it appears as a strictly local operator. Then the analysis follows strictly that of the previous 
section identifying the relevant normalization parameters with the 
zero momentum value of the vertices whose dimension does not 
exceed 4 taking into account also the dimension of the source. 

 We can therefore apply our results to multilocal products of 
operators to study their Wilson short-distance expansion.  
We can therefore apply our results to
To put into evidence the origin and the nature of this expansion
we simplify as much as possible our discussion referring to the 
product of two field operators, that is: to the operator whose bare structure
is the product of two bare fields at different points ($\f(x)\f(0)$). 
The crucial new aspect of our analysis consists in considering this 
product, for what concerns the effective lagrangian, as a single 
composite operator. In particular we associate a source $\u_X$ to it and
we study the dependence of $\ll$ on this source.

Notice that considering the product of two fields in different 
points as a single composite operator implies a modification of the 
concept of connectedness of a diagram.

As discussed above, the first step in the analysis of our composite
operator consists in assigning a dimension to $\u_X$.
Owing to the fact that the contributions of Feynman diagrams to the
tree vertices involving our operator depend on
$x$ through a phase factor $e^{ipx}$, where p is a partial sum of the external
momenta, we have that the largest possible value of the dimension of $\u_X$
is $2$, the same the local operator $\f^2$. 

Having specified the dimension of the source of our bilocal operator, we 
consider now its normalization conditions. According to the 
above discussion in order to
define $(\f (x)\f (0))$ 
as a single composite operator, we have to assign its vacuum 
vertex, that is: the coefficient of $\ll$ linear in the source $\u_X$
and of degree zero in the field, and the vertex with two legs with vanishing
momentum. These vertices can be easily deduced from the vertices of $\ll$
involving only $\f$-legs. It is sufficient to consider respectively the 
vertices $\c^{(2,0)}(\v p)$ and $\c^{(4,0)}(\v p,-\v p,0,0)$, to multiply 
then by $e^{i\v p\cdot\v x}\D(p)^2$ and to integrate with respect to 
$\v p$.

It is therefore possible, in our theory, to define for 
every $x$ a renormalized operator of dimension $2$ corresponding to the product
of two bare fields at different points.  This is identified with an
operator of dimension two satisfying the following normalization
conditions:

i) For the vacuum vertex, :
\be \int {d^4k\o (2\pi)^4}{\d\o\d\u_X (\v k)}\ll\vert_{\f=\u_X=0}=
\int {d^4p\o (2\pi)^4}e^{ipx}\la_R^2\D (p)
\[1+\c^{(2,0)}(p^2)\D (p)\]\ .\label{5,1}\ee

ii) For the coefficient with two field legs at zero momentum:
\be \int {d^4k\o (2\pi)^4}{\d\o\d\u_X (\v k)} {\d^2\o\d\f^2(\v k)} \ll
\vert_{\f=\u_X=0}=
\int {d^4p\o (2\pi)^4}e^{ipx} \D^2 (p) \c^{(4,0)}(\v p,-\v p,\v 0,\v 0)
\ .\label{5,2}\ee

As it is evident already at the tree level, these normalization conditions
are by no means uniform in $x$, in particular the vacuum vertex is proportional
to ${1\o x^2}$ at short distances. This lack of uniformity in $x$ can however
be taken into account in the following way. We consider the
linear combination of composite operators:
\be N_2^{(\la_R)}[\f(x)\f(0)]\equiv \f(x)\f(0)-A(x)-
B(x)\ N_2^{(\la_R)}[\f^2(0)]\ .
\label{5,3}\ee Here $A(x)$ and $B(x)$ are c-numbers chosen in such a way 
that the new composite operator 
$N_2^{(\la_R)}[\f(x)\f(0)]$, satisfies normalization conditions uniform in $x$.
More precisely:

i) Its vacuum vertex has to vanish.
 
ii) The two field vertex at zero momentum has to be equal to two.

Now, remembering the analysis of the evolution equations, 
we notice that the new operator has vertices uniformly bounded in $x$
owing to the uniformity of the normalization conditions at $\la_R$.
Therefore the introduction of $N_2^{(\la_R)}[\f(x)\f(0)]$ can be seen as a
decomposition of a singular operator product into the sum of a singular and
a regular part:
\be \f(x)\f(0)=A(x)+B(x)\ N_2^{(\la_R)}[\f^2(0)]+N_2^{(\la_R)}[\f(x)\f(0)]\ .
\label{5,4}\ee
The first two terms in the right-hand side give the singular part. Indeed
taking into account (\ref{5,1}) and (\ref{5,2}) it is 
possible to verify that:
\be A(x)\sim{1\o x^2}P\(ln(x\la_R)\)\ ,\label{5,5}\ee
and that:
\be B(x)\sim Q\(ln(x\la_R)\)\ ,\label{5,6}\ee
where $P$ and $Q$ are polynomials of increasing degree with the perturbative
order. This requires a little work on the evolution equation (\ref{4,18}) 
that,
after UV limit, can be transformed into the Callan-Symanzik equation [7]
and then used to study the presence of log's.

The decomposition (\ref{5,4}) is the essential part of Wilson operator product
expansion \cite{3} , it can be obviously extended to multilocal products of 
composite operators.
Starting from this expansion it is possible to analyze 
the Green functions of the theory \cite{4} as distributions
recovering the results of Bogoliubov,
Parashiuk, Hepp, Zimmermann, Epstein, Glaser and Stora and hence proving that
our theory leads to the same Green functions as the subtraction methods
mentioned in the introduction.

\vfill\eject
\salto
\sec{The quantum action principle.}
\salto

Now we discuss how one can characterize the invariance of a theory 
under a system of, in general non-linear, continuous field 
transformations.

In the Feynman functional framework, the invariance of the theory 
corresponds to that of the functional measure and its consequences
are simply exhibited by introducing the field transformations as a
change of variables in (\ref{2,30}).

A simple but significant example of a model with non-linear symmetry 
is provided by the 
two-dimensional non-linear $\sigma$-model. With a simple choice of 
coordinates the classical action of this model is written:
\be S=\int d^2x\[z\[(\p\v{\pi})^2+\(\p\sqrt{F^2-\pi^2}\)^2\]+
m^2\(\sqrt{F^2-\pi^2}-F\)\]\ ,\label{6,1}\ee
where $\v{\pi}$ is an isovector field with vanishing dimension.
This action is invariant under the linear field transformations 
corresponding to rotations in isotopic space. However to characterize 
completely (\ref{6,1}) we have to exploit the transformation property of 
the action under the non-linear transformation:
\be \v{\pi} \to\v{\pi}+ \v{\epsilon}\sqrt{F^2-\pi^2}\ ,\label{6,2}\ee
where $\v{\epsilon}$ is an infinitesimal, space independent isovector.
It is easy to verify that 
the corresponding variation of the classical action is:
\be -\int d^2x\ m^2\ \v{\epsilon} \cdot \v{\pi} \ .\label{6,3}\ee
If one asks, 
together with the linear isotopic invariance, that the 
variation of the action under (\ref{6,2}) be (\ref{6,3}),  one identifies 
the action (\ref{6,1}) up to the choice of $z$. This example is presented 
to put into evidence the role of symmetries in the identification of the 
action and hence in the construction of quantum field theories.

In the general case we shall consider a multicomponent field $\f^i$ and a set
of infinitesimal, non-linear, rigid
field transformations, that we shall write, disregarding the indices, according:
\be \f_{\v p}\to\f_{\v p}
+\epsilon\[ T\f_{\v p} + k\({p\o\lao}\)  
P_{0,\v p}(\ko,\u,0)\]\ ,\label{6,4}\ee where the matrix $T$ is assumed 
to be traceless and to satisfy:
\be T C(p) \pm C(p) T^T=0\ ,\label{ct}\ee
where the sign has to be chosen in agreement with the field statistics.
It is apparent that this equation
 is equivalent to the invariance of the free part of the action under
the linear part of the field transformation.
The polynomial $P_{0,\v p}(\ko,\u,0)$ accounts for the strictly non-linear 
part of the transformation, let us indicate by $d_P$ the canonical dimension
of the corresponding operator.
 
To define the composite operator $P$ within the functional framework,  
we introduce its source $\c$, with dimension $4-d_P$, by the substitution:
\be L_0(\ko,\u)\to L_0(\ko,\u)-\sum_{\v p}\c_{\v p}\, P_{0,-\v p}(\ko,\u)+O(\c^2)
\equiv L_0(\ko,\u,\c)\ .\label{6,5}\ee

Notice that the introduction of a composite operator of source $\c$ is
in general accompanied by terms of degree higher than one in $\c$
which appear in the term $O(\c^2)$ in (\ref{6,5}). The 
presence of these terms induces
a change of the infinitesimal transformation law (\ref{6,4}) that
we shall write
according:
\bea \f_{\v p}\to\f_{\v p}+\epsilon\[ T\f_{\v p}
-k\({p\o\lao}\)  
\p_{\c_{-\v p}} L_0(\ko,\u,\c)\]\equiv\nn  
\f_{\v p}+ \epsilon\[ T\f_{\v p}
+k\({p\o\lao}\)
P_{0,\v p}(\ko,\u,\c)\]\ .\label{6,6}\eea
In general the substitution shown in (\ref{6,5}) produces symmetry breaking 
terms of $O(\c )$. We shall see however that in the case of gauge theories
the nilpotency of the transformation (\ref{6,4}) allows a precise control of 
these $O(\c )$ terms.

Introducing (\ref{6,6}) as a change of variables into the Feynman formula 
(2,30) 
and requiring the invariance of the functional integral, we get, to first 
order in $\epsilon$:
\bea \in\SP \[k\({p\o\lao}\)  
\[\(j_{\v p}-\f_{\v p}C(p)-\p_{\f_{-\v p}}L_0\)
\p_{\c_{\v p}} L_0
+ \p_{\f_{-\v p}}\p_{\c_{\v p}} L_0\] \right.\nn\left.
- j_{\v p}\,T \f_{-\v p} +\f_{\v p}\,T^T \p_{\f_{\v p}}L_0 \]
e^{-S}
=0\ .\label{6,7}\eea

Let us now introduce the  composite operator:
\bea \D\equiv\sum_{\v p}\[k\({p\o\lao}\) 
 \[\(\f_{\v p}\, C(p)+\p_{\f_{-\v p}}L_0\)
\p_{\c_{\v p}} L_0
- \p_{\f_{-\v p}}\p_{\c_{\v p}} L_0\] \right.\nn\left. 
- \f_{\v p}\,T^T \p_{\f_{\v p}}L_0 \]
\ ,\label{6,9}\eea or, using the simplified notation introduced in section (2):
\bea\D\equiv \SP\[\kpo\f_{\v p}\( C(p)\pgp L_0-T^T\pp L_0\)\right.\nn\left.+
k^2\({p\o\lao}\)\(\pgmp L_0\pp L_0 -\pgmp\pp L_0\)\]\ ,
\eea
and 
define
\be \D\cdot Z\equiv N\in\,\D\, e^{-S}\label{6,10}\ee the generator of 
the Schwinger functions with the 
insertion of the operator $\D$. This is computed in the functional 
framework adding to the bare action $\epsilon\D$ 
and taking the derivative of $Z$ with respect to $\epsilon$ at
$\epsilon =0$. 
Remembering that the support of  $j$ is contained in a 
sphere of radius $\la_R$, which is smaller than $\lao$, we can
disregard the cut-off factor $k\({p\o\lao}\)$ in the first term 
in the left-hand
side of (\ref{6,7}) which can be written:
\be \sum_{\v p} j_{\v p}\(\p_{\c_{\v p}} -T \p_{j_{\v p}}\)Z=
\D\cdot Z\ .\label{6,11}\ee
Now, if the couplings of the source $\c$ 
are suitably normalized and therefore
the theory has UV limit in the presence of $\c$ (the operator $P$ is 
finite), from (\ref{6,11}) we trivially see that the operator $\D$ is finite.

Equation (\ref{6,11}) is a broken Ward identity, it is
 sometimes referred to as "the Quantum Action Principle". 
 For an introduction to its
consequences in a regularization independent framework we refer to 
\cite{4},\cite{5}. 

Of course, the finiteness of $\D$ is interesting since it ensures the 
meaningfulness of the broken Ward identity (\ref{6,11}). However this is not
the final goal of our study which is devoted to the construction of 
quantum theories with symmetry properties. Therefore the equation that we 
have to discuss and possibly to solve is:
\be \D=O(\c )\ .\label{6,12}\ee This is a system of equations for the bare, or 
 better, the normalization constants of the theory that should allow to 
identify the invariant theory. The general idea is that the symmetry 
breaking effect induced by the regularization of the Feynman integral 
should be compensated fine-tuning non-invariant terms in the bare action.
This is why we shall call (\ref{6,12}) the fine-tuning equation.

Until now we have  considered on equal footing the possibility 
of defining
the left-hand side of (\ref{6,12})  before and after the UV limit
 provided that the theory
be regular in this limit.
The first point 
to be clarified in our analysis of this equation is to decide 
if we 
are going to study it before or after UV limit. In the case in which
 the existence 
of the UV limit is guaranteed by simple
power counting arguments,  as it is in the framework of perturbation
theory, it is natural to consider the situation after
this limit. In this case one has to study, instead of (\ref{6,11}),
the corresponding equation for
the effective action since the bare action is  UV-singular.

Let us then discuss the fine-tuning equation after the UV limit.
To extend our results to the effective 
theory we repeat the previous analysis referring to the effective lagrangian and  
we introduce the effective breaking:
\bea \De \equiv \SP\[\kp\f_{\v p}\( C(p)\pgp \L-T^T\pp \L\)\right.\nn\left.+
\kdp\(\pgmp \L\pp\, \L -\pgmp\pp\, \L\)\]\ ,\label{def2}
\eea
Then (\ref{6,12}) 
is naturally replaced with:\be \D_{eff} =O(\c )\ .\label{6,15}\ee

This fine-tuning condition (\ref{6,15}) is equivalent to an infinite 
number of equations for $\De$. However it is apparent from (\ref{6,9})
that $\D$ has limited dimension provided the field transformations 
(\ref{6,6}) change the field dimension by a limited amount. More precisely,
in the limit $\lao\to\infty$, 
the bare operator $\D$ becomes a local operator of dimension $3+d_P$. 

Therefore, according to the remark at the beginning of section 5 
concerning composite operators, $\D$ is completely equivalent to a strictly local
operator whose dimension is equal to 4 plus the increase 
of field dimension in the transformations (\ref{6,4}). 

Furthermore, 
computing explicitly the $\la$-derivative of the right-hand side of 
(\ref{def2}) and taking into account (\ref{3,2}) it can be seen 
that $\De$ satisfies, before the UV limit, the evolution equation:
\be \la\p_{\la}\De=M\[\De\]\ ,\label{ev}\ee
if the theory is regular in the UV limit (\ref{ev} ) is expected 
to hold true also after this limit. Therefore we can conclude that
$\De$ is determined by its normalization conditions and hence the vanishing
condition for $\De$ is equivalent to a finite number of equations concerning the
normalization conditions for $\De$. 

Let us notice that  (\ref{ev} ) 
can be verified directly using the evolution equation for
$\L$. Indeed one can compute the left-hand side of (\ref{ev} ) according:
\bea\ldl\De=\SQ\ldl\kdq \(\pgq\L\pmq\L-\pgq\pmq\L\)\nn
\SP\SQ\kdp{\ldl\kdq\o C(q)}\[\pgp\pq\L\pmq\L\pmp\L\acca
+\pgp\L\pmq\L\pmp\pmq\L-{1\o2}\(\pgp\pq\pmq\L\pmp\L+\pq\pmq\pmp\L\pgp\L\)
\acca -\pmp\pgp\pq\L\pmq\L-\pq\pgp\L\pmp\pmq\L +{1\o2}\pmp\pgp\pmq\pq\L\]
\nn
\SP\SQ\kp{\ldl\kdq\o C(q)}\[\fp\, C(p)\(\pgp\pq\L\pmq\L-{1\o2}\pgp\pq\pmq\L
\)\acca -\fp T^T\(\pp\pq\L\pmq\L-{1\o2}\pp\pq\pmq\L\)\]\ ,\label{ldld}
\eea 
and, taking into account the definition of $M$ given in (\ref{3,13}),
 the right-hand side is computed according:
\bea M\[\De\]=\SP\SQ\kdp{\ldl\kdq\o C(q)}\[\pgp\pq\L\pmp\L\pmq\L\acca
+\pgp\L\pmp\pq\L\pmq\L-\pgp\pmp\pq\L\pmq\L-\pgp\pmq\L\pmp\pq\L\acca-{1\o2}\(
\pq\pmq\pgp\L\pmp\L+\pq\pmq\pmp\L\pgp\L-\pgp\pmp\pq\pmq\L\)\]\nn
+\SP\SQ\kp{\ldl\kdq\o C(q)}\[\fp C(p)\(\pgp\pq\L\pmq\L-\pgp\pq\pmq\L\)\acca
-\fp T^T\(\pp\pmq\L\pq\L+\pp\pq\pmq\L\)\]\nn
\SQ{\ldl\kdq\o C(q)}\[C(q)\(\pgq\L\pmq\L-\pgq\pmq\L\)\acca
-T\(\pq\L\pmq\L-\pq\pmq\L\)\]\ \label{M}.
\eea
In (\ref{M}) we have left  to the reader the bookkeeping of the field indices.
If this is correctly made it is apparent that the last line of (\ref{M})
vanishes due to (\ref{ct}), in which one has to replace $C$ with $C^{-1}$, and
that (\ref{ldld}) and (\ref{M}) coincide, thus confirming (\ref{ev} ).

Therefore, after the general discussion concluding section  
4, we can say that (\ref{6,15}) is guaranteed in the UV limit if $\De$ 
satisfies vanishing normalization conditions.

At this point of the analysis it is convenient to perform the infinite 
volume limit. As discussed in section 3 this limit is completely harmless,
at least in perturbation theory. Therefore, from now on, we understand
that all the sums over the momenta in the previous formulae are replaced
with the corresponding integrals.

Now we can express the vanishing of these normalization conditions in a simple 
form introducing ${\cal T}_d F$: 
the local approximant of dimension $d$ to a translation 
invariant functional $F$.
This approximant is defined as the integrated local functional of dimension
$d$ of fields and 
sources whose vertices coincide with those generated by $F$ 
up to terms with  momentum degree equal to $d$ minus the total dimension of the fields and 
sources corresponding to the external legs of the vertex. Here, as usual,
 we assign the source of a local operator the complement to four of the dimension of the 
operator.
For example if $F$ depends only on $\f_i$ we have:
\bea{\cal T}_3 F=F[0] +\int d^4x\f_i (x)\[{\d\o\d\f_i(0)} F+{1\o2}\f_j(x)
{\d\o\d\tilde\f_j(0)}{\d\o\d\f_i(0)}F\r.\nn\l. +{i\o2}
\p_{\mu}\f_j(x)\p_{p_{\mu}}{\d\o\d\tilde\f_j(0)}{\d\o\d\f_i(0)}F+\r.\nn\l. 
{1\o6}\f_j (x)\f_k(x){\d^2\o\d\tilde\f_j(0)\d\tilde\f_k(0)}{\d\o\d\f_i(0)}F\]
\vert_{\f=0}\ 
,\label{al}\eea
where we have set 
\be\f(x) =\int d^4pe^{-ipx}\t \f (p)\ .\ee
Given a local operator $O$ with source $\eta$ and dimension $d_O$, this 
definition automatically induces that of 
the effective local approximant of dimension
$d$ to $O$, according:
\be {\cal T}_dO_{eff}\equiv{\cal T}_d{\d \L\o\d\eta}\ .\label{leap}\ee
In perturbation theory the effetive local approximants 
to a local operator have a very 
special property.
Let us introduce the perturbative order $\u_O$ of a 
local operator, as the first order
in $\hbar$ at which the operator contributes to the Green functional generator $Z$.
In general this order appears as the $\hbar$-power of  a factor in front of the
 bare operator.

The above mentioned special property is the following: all the effective local
approximants to $O$ with dimension higher or equal to $d_O$
coincide up to terms of order $\hbar^{\u_O+1}$
That is:\be {\rm if}\quad\, d\geq d_O\quad\rightarrow\quad {\cal T}_dO_{eff}=
{\cal T}_{d_O} O_{eff}+O\(\hbar^{\u_O+1}\)\ .\label{dind}\ee

This property is easily understood recalling the interpretation of $\L$, given at the 
beginning of section 4, as the functional generator of the connected and amputated
Feynman amplitudes corresponding to the propagator:
\be {k^2\({p\o\lao}\)-k^2\({p\o\la}\)\o p^2+m^2}\ ,\ee
and the vertices correspnding to $L_0$. Then the contributions
 of order $\hbar^{\u_O}$ to the vertices generated by
$O_{eff}$ are given by the tree approximation connected and
 amputated amplitudes and the same holds true for their momentum derivatives.
Owing to the fact that the above propagator vanished with all its momentum 
derivatives for vanishing momentum we see that the only contribution of order
$\hbar^{\u_O}$ to ${\cal T}_dO_{eff}$ come from connected
and amputated tree diagrams without internal lines, that is those coinciding
 with a single vertex. These vertices are determined by the bare
structure of the operator and their local approximant 
do not depend on $d$ provided $d\geq d_O$. 
This proves the mentioned property.

In the case of ${\cal T}_{d_{\De}}\De$ and of ${\cal T}_4\S$ we shall,
from now on, understand the dimension of the local approximant.

Since the couplings appearing in $\Dl$ are in one-to-one correspondence 
with the normalization parameters of $\De$ it is now clear that 
if the theory is finite in the UV limit, the fine tuning equation
 (\ref{6,15}) is equivalent to \be \Dl=O(\c )\ ,\label{s}\ee
that is to:
\bea{\cal T}\ipp\[ \f (p)C(p){\d\L\o\d\c (p)}+{\d\L\o\d\f (p)}{\d\L\o\d\c (-p)}
\acca - \f (p)T^T{\d\L\o\d\f (p)}-\hbar\kp{\d^2\L\o\d\f (p)\d\c (-p)}\]=O(\c )\ ,
\label{appr}
 \eea where we have taken into account that $k(0)=1$ and that all the
partial derivatives of $k$ vanish at the origin. We have also considered that
$\L$ is a functional of $\kp\f (p)$ and $\c (p)$. Notice that in (\ref{appr})
we have introduced in the last term the $\hbar$ coefficient that appears 
in a loop-orderd perturbation theory in which one introduces $\hbar$ as 
loop counting parameter. 

For future purposes it is convenient to introduce here two further fuctionals:
\be \S\equiv\ipp\[{\f (-p) C(p)\f (p)\o2}-\c (-p) T\f (p) +\L\]\ ,\label{efac}
\ee
and
\bea\hDe\equiv -\hbar^2e^{\S\o\hbar}\ipp\,\kp{\d^2\o\d\f (p)\d\c (-p)}
 e^{-\S\o\hbar}\nn
\ipp\,\kp\[ {\d\S\o\d\f (p)}{\d\S\o\d\c (-p)}
-\hbar{\d^2\S\o\d\f (p)\d\c (-p)}\]\nn
=\ipp\,\kp\[\({\d\L\o\d\c (-p)}-\f (p)T^T\)\(C(p)\f(-p)+{\d\L\o\d\c (p)}\)\acca
-\hbar{\d^2\S\o\d\f (p)\d\c (-p)}\] +O'(\c)\ ,\label{hde}
\eea
It is apparent that:\be\Dl =\hDl +O'(\c)\ ,\ee and hence the fine tuning equation
can be written and discussed in terms of $\hDe$ instead of $\De$.
We shall present the further developments of our method in the next section
 applying them directly to a non-abelian gauge theory.

Beyond the perturbative level the existence of an UV limit is
far from being guaranteed. However there is some evidence of 
a rather general connection between the existence of this limit and 
asymptotic freedom \cite{9}. In this case the study of (\ref{6,12}) 
or (\ref{6,15})
must precede the UV limit since the theory with broken symmetry in general
is not expected to be asymptotically free.

Notice however that in the presence of an UV cut-off $\lao$ the wanted 
symmetry can be unreachable at the bare level. This happens typically
in a gauge theory in which (\ref{6,12}) is
unsolvable for finite UV cut-off. 

Therefore at first sight the construction of a fully quantized gauge theory 
seems impossible. However in the framework of 
the effective theory it might be possible
to fine-tune the low energy parameters 
under the condition that $\D_{eff}$ given in 
(\ref{def2}) 
satisfy normalization conditions vanishing with ${\la\o\lao}$ . 

If this condition is met, considering the solution of
the evolution equations discussed in section 4, we see that the breaking
is irrelevant and hence, if the theory has a regular
UV limit, $\D$ vanishes in this limit at least as ${\la\o\lao}$. To satisfy 
this last condition 
one should ask the fine-tuned theory to be asymptotically free. Postponing the 
dream of the non-perturbative construction of a gauge theory to better 
times, we shall limit our present discussion to the perturbative situation.

\salto
\sec{Analysis of the $SU(2)$ Yang-Mills model}
\salto
To proceed further 
with the analysis of (\ref{6,15}) it is convenient to refer to the 
highly non-trivial example of a pure $SU(2)$ gauge theory, whose field content
consists of the gauge field $A^a_{\mu}$, of the Lagrange multiplier $b^a$
and of the anticommuting Faddeev-Popov ghosts $c^a$ and $\bar c^a$. All the
fields are isotopic vectors. In terms of the gauge field strength:
\be G^a_{\mu\nu} = \p_{\mu} A^a_{\nu} - \p_{\nu} A^a_{\mu}- g\epsilon^{abc}
A^b_{\mu} A^c_{\nu}\ ,\label{6,17}\ee where $\epsilon^{abc}$ is the 
antisymmetric
Ricci tensor, and of the covariant derivative:
\be D_{\mu} c^a = \p_{\mu} c^a -g\epsilon^{abc} A^b_{\mu} c^c\ ,\label{6,18}\ee
the "classical action" of the theory is given:
\be S_{cl} = \int d^4x\ \[{z\o4} G^a_{\mu\nu} G^a_{\mu\nu} + {b^a b^a \o2}
+ i b^a \p_{\mu} A^a_{\mu} +\p_{\mu} \bar c^a D_{\mu} c^a\]\ .\label{6,19}\ee
From this equation one can easily deduce the covariance matrix of the fields.
It is also possible to single out the symmetries characterizing the theory at
the quantum level. It turns out that, given the $b^a$-dependent part of 
(\ref{6,19}),
$S_{cl}$ is identified by the invariance condition under the non-linear, 
nilpotent field transformations:
\be A^a_{\mu} \rightarrow A^a_{\mu}  +\eta D_{\mu} c^a\ ,\label{6,20}\ee
\be c^a \rightarrow c^a + \eta {g\o2}\epsilon^{abc} c^b c^c\ ,\label{6,21}\ee
\be \bar c^a \rightarrow \bar c^a + i \eta b^a\ ,\label{6,22}\ee
\be b^a \rightarrow b^a \ .\label{6,23}\ee
Notice that here $\eta$ is an anticommuting constant that replaces 
$\epsilon$ in the definition of $\D$.

A further necessary comment concerning (\ref{6,19}), 
is that the field $b^a$ plays 
the role of an auxiliary field since the action is at most quadratic in it
and hence it does not appear into the interaction lagrangian. It follows that
the Wilson renormalization group starting from a bare action with this 
property will generate a $b$-field independent effective lagrangian.

Owing to the anticommuting character of the transformations (\ref{6,20}-
\ref{6,23}), 
quantizing the theory, we have to introduce an anticommuting vector-isovector
source $\c^a_{\mu}$ of dimension two, coupled to
 $-g\c_{\mu}^a \epsilon^{abc} A^b_{\mu} c^c$ and a 
commuting isovector $\zeta^a$ of dimension two, coupled to the variation of
$c^a$. Adding to the action these source terms together with the sources
of the fields we have:
\bea S_{cl} = \int d^4x\ \[{z\o4} G^a_{\mu\nu} G^a_{\mu\nu} + \a{b^a b^a \o2}
+ i b^a \p_{\mu} A^a_{\mu} +\p_{\mu} \bar c^a D_{\mu} c^a
\right.\nn  \left. -g\c_{\mu}^a \epsilon^{abc} A^b_{\mu} c^c -
\zeta^a {g\epsilon^{abc}\o2}c^bc^c-j_{\mu}^a A^a_{\mu} -J^a b^A
-\xi^a \bar c^a -\bar\xi^a c^a\]\ .\label{6,24}\eea
Notice that $\xi^a$ and $\bar\xi^a$ are anticommuting sources and that 
there is a conserved ghost charge that vanishes for the vector field, is 
equal to $+1$ for the $c$ field, $-1$ for the $\bar c$ and $\c$ source and 
$-2$ for the $\zeta$ source. 

In order to quantize this theory we introduce a cut-off in much the 
same way as for the scalar field, choosing the same function $k$ for all the
field components. In this way we maintain at the quantum level
the invariance of the theory under rigid isotopic symmetry. however this
automatically introduces an unavoidable breaking into the Ward identity 
corresponding to (\ref{6,20}-\ref{6,23}). Indeed, 
considering the cut-off transformations:
\be A^a_{\mu \v p}\rightarrow A^a_{\mu \v p}+\eta\(\kpo \dg L_0 + 
i p_{\mu} c^a_{-\v p}\)\ ,\label{6,26}\ee
\be c^a_{ \v p}\rightarrow c^a_{ \v p} +\eta \kpo\dz L_0\ ,\label{6,27}\ee
\be \bar c^a_{ \v p}\rightarrow \bar c^a_{ \v p} +i\eta b^a_{\v p}\ 
,\label{6,28}\ee
and following the analysis described above, we get the broken Ward identity:
\be \SP \(j^a_{\mu \v p}\(\dg +ip_{\mu}\p_{ \xi^a_{\v p}}\)-
\bar\xi^a_{\v p}\dz-i\xi^a_{\v p}\p_{J^a_{\v p}}\)
Z\equiv{\cal S} Z=\D\cdot Z\ ,\label{6,28b}\ee where:
\bea \D=\nn\SP\[\((\d_{\mu\nu}p^2-p_{\mu}p_{\nu}) A^a_{\nu \v p} 
+ p_{\mu} b^a_{\v p} 
+\da L_0\)\(\kpo \dg L_0 + i p_{\mu} c^a_{-\v p}\)-
\right.\nn  \left.  \(p^2 \bar c^a_{\v p} - \dc L_0\) \kpo\dz L_0
-i\( p^2 c^a_{\v p} +\p_{\bar c^a_{-\v p}} L_0\) b^a_{-\v p} 
-\right.\nn  \left. 
\kpo\(\da\dg +\dc\dz\)L_0\]\ ,\label{6,29}\eea
Substituting the classical lagrangian into this equation and disregarding 
the cut-off factors, one finds that at the classical limit $\D$ is 
equal to:
\be \D_{cl}=-ig\epsilon^{abc}\SP\SQ p_{\mu} \c^a_{\mu\v p}c^b_{\v q}
c^a_{-\v q-\v p}=-i\SP p_{\mu}\dz S_{cl}\ .\label{dcl}\ee

For a generic choice of the bare action it is apparent that this breaking 
is $b$-dependent. Now this is a true disaster since, even if the breaking
were compensable by a suitable modification of the action, this would introduce
$b$-dependent interaction terms. However this difficulty can be avoided
if one chooses the bare action so that:
\bea -i p_{\mu}\dg L_0=\p_{K_{\lao}\bar c^a_{\v p}}L_0\ .\label{6,30}\eea
Indeed substituting (\ref{6,30}) into (\ref{6,29}) 
it is apparent that the $b$-dependent terms
in $\D$ annul each other.

Now, assuming that the theory remain finite in the UV limit, we translate 
the above results in terms of the effective theory in the infinite  volume 
limit.
 
We have in particular that the constraint (\ref{6,30}) remains true if we 
replace in them $L_0$ with $\L$. That is
\bea -i p_{\mu}\dg \L=\p_{K_{\la}\bar c^a_{\v p}}\L\ .\label{6,30'}\eea
Indeed it is apparent that both sides  of (\ref{6,30'}) satisfy the same 
linear evolution equation. In particular 
for the left-hand side this equation reads: 
\be \la\p_{\la}\dg \L=M\[\dg\L\]\ .\ee
Taking into account (\ref{dcl}) we assume for $\De$ the following fine 
tuning condition:
\be\Dl =-i\ipp p_{\mu}\c^a_{\mu}(p){\d{\cal T}\L\o\d\zeta^a(p)}\ .\label{ftc}\ee

Now, following the line shown in the previous section we introduce
 the effective action:
\bea \S[\f,\c] 
=\IP \[\t A^a_{\mu}(-p){\d_{\mu\nu}p^2-p_{\mu}p_{\nu}\o2}\t A^a_{\nu}(p)
+\bar{\t c}^a(-p) p^2\t c^a(p)-\r.\nn\l.
i\t \c^a_{\mu}(-p)p^{\mu}\t c^a(p)\] 
+\L\ ,\label{6,32}\eea
and the modified breaking $\hDe$ defined in (\ref{hde}) that in the
present case is written:
in terms of it:
\bea \hDe= -\hbar^2e^{\S\o\hbar}\IP\kp\(\dea\deg+\dec\dez\)e^{-\S\o\hbar} 
=\nn\IP\kp\[\dea \S\deg\S+\dec \S\dez\S-\r.\nn\l.
 \hbar
\(\dea\deg+\dec\dez\)\S\] \ .\label{6,33}\eea

Comparing the true breaking (\ref{def2}) with the modified one given above
we see that:
\be\hDl =\Dl+i\ipp p_{\mu}\c^a_{\mu}(p){\d{\cal T}\L\o\d\zeta^a(p)}\ ,\ee
and hence we are left with the homogeneous fine-tuning condition:
\bea \hDl=\nn-\hbar^2
{\cal T}_5\[e^{\S\o\hbar}\IP\kp\(\dea\deg+\dec\dez\)e^{-\S\o\hbar}\] 
=\nn\IP\[{\cal T}_5\(\dea \S\deg\S+\dec \S\dez\S\)-\r.\nn\l.
 \hbar{\cal T}_5\kp
\(\dea\deg+\dec\dez\)\S\]=0 \ .\label{6,33b}\eea
Notice that the cut-off factor in the first two terms under integral has 
disappeared since, by momentum conservation, the ${\cal T}$ operator 
restricts the $p$ variable corresponding to these terms to zero.

It is important to notice here that, as it will be apparent in a moment,
we can choose this homogeneous condition due to the nilpotency of the
 classical transformations (\ref{6,20}-\ref{6,23}).

In the present case the 
differential operator 
\bea \IP\kp {\d\o\d\f (-p)}{\d\o\d\c (p)}\equiv\nn
\IP\kp\(\dea\deg+\dec\dez\)\ ,\label{6,34}\eea
is nilpotent owing to the anticommuting nature of the ghost
field $c^a$ and of the source $\c^a_{\mu}$.

This nilpotency of the differential operator (\ref{6,34}) leads 
to the following equation for $\hDe$ :
\bea -\hbar\IP\kp {\d\o\d\f (-p)}{\d\o\d\c (p)}\hDe 
e^{-\S\o\hbar} =\nn\IP\kp\[ {\d\o\d\f (-p)}\S{\d\o\d\c (p)}\hDe+
 {\d\o\d\f (-p)}\hDe{\d\o\d\c (p)}\S-\r.\nn\l.\hbar
 {\d\o\d\f (-p)}{\d\o\d\c (p)}\hDe\]
e^{-\S\o\hbar} =0\ ,\label{6,35}\eea where we have taken into account 
(\ref{6,33}) 
and the anticommuting character of $\hDe$.

The last identity, that in the gauge case is written:
\bea \IP\kp\[\dea \S\deg\hDe+\deg \S\dea\hDe+\r.\nn\l.\dec \S\dez\hDe+  
\dez \S\dec\hDe-\r.\nn\l.
 \hbar
\(\dea\deg+\dec\dez\)\hDe\]=0\ ,\label{6,36}\eea 
plays the role of a consistency 
condition for the breaking analogous to the Wess-Zumino \cite{10} 
consistency condition
for the chiral anomaly. 

Now we show how in the perturbative case this consistency condition 
turns out to be sufficient to prove the solvability of the fine-tuning 
equation (\ref{6,15}). 

As repeated many times in the previous sections, perturbation theory is 
built by a recursive procedure in $\hbar$. Hence we have to apply this 
procedure to the fine-tuning equation. Therefore we develop the effective 
action $\S$ and the effective breaking $\hDe$ in power series of $\hbar$ and 
we label the $k^{th}$-order terms of these series by the index $k$.

Now, if we assume the tree approximation normalization condition
\be{\cal T}\S_0=
\int d^4x\ \[{1\o4} G^a_{\mu\nu} G^a_{\mu\nu} 
+\(\p_{\mu} \bar c^a -\c_{\mu}^a \)D_{\mu} c^a
-\zeta^a {g\epsilon^{abc}\o2}c^bc^c\]\ ,\label{6,37}\ee
 we have
\bea  \hDl_0=
\IP\(\dea \Sl_0\deg\Sl_0+
\r.\nn\l.\dec \Sl_0\dez\Sl_0\)=0 \ .\label{6,33c}\eea
Indeed one can directly verify using the fact that $\S_0$ is 
invariant under the transformations (\ref{6,20}-\ref{6,21}).

Assuming that
\be \hDl_{k}=0\quad{\rm for}\quad k<n\ ,\label{6,38}\ee 
it follows that $\hDe_{k}=0$ and 
hence the consistency condition
\bea \IP\kp\[\dea \S_0\deg\hDe_n+\deg \S_0\dea\hDe_n+\r.\nn\l.\dec 
\S_0\dez\hDe_n+  
\dez \S_0\dec\hDe_n\]=0\ .\label{6,39}\eea

Selecting the local approximant of dimension 6 to (\ref{6,39}) and taking 
into account (\ref{6,37}) and (\ref{6,40}) we get the perturbative 
consistency condition:
\bea \IP\[\dea \Sl_0\deg\hDl_n+\deg \Sl_0\dea\hDl_n+\r.\nn\l.
\dec \Sl_0\dez\hDl_n+  
\dez \Sl_0\dec\hDl_n\]\equiv\nn
{\cal D}\hDl_n=0\ .\label{6,41}\eea 
Here we have used (\ref{dind}). From (\ref{6,37}) we have the explicit form of 
${\cal D}$:
\bea {\cal D}=-\int d^4x\[D_{\mu} c^a{\d\o\d A_{\mu}^a} -{g\epsilon^{abc}\o2}c^bc^c
{\d\o\d c^a}+\r.\nn\l.
\(D_{\nu} G^a_{\nu\mu}+g\epsilon^{abc}\c_{\mu}^c c^c\){\d\o\d \c_{\mu}^a}
+D_{\mu}\c_{\mu}^a{\d\o\d \zeta^a}\]\ .\label{dd}\eea

It is important 
to notice here that the differential operator ${\cal D}$ defined above
is nilpotent. Therefore any element of the image of ${\cal D}$ solves 
(\ref{6,41}).
 
At this point of our analysis it is convenient to write explicitly the 
local approximant of dimension 5 to $\hDe_n$. From 
now on we shall put $\bar c=0$ without any loss of 
generality since the dependence on this field is strictly connected to that 
on the source $\c^a_{\mu}$. Owing to the euclidean, 
isotopic $(SU(2))$ and translation invariance and recalling 
that $\hDe$ has ghost charge  $+1$, we have:
\bea\hDl_n=\int d^4x\[a_n\ep\c_{\mu}^ac^b \p_{\mu} c^c+b_n
\c_{\mu}^aA_{\mu}^bc^ac^b+\r.\nn\l. c_nA_{\mu}^a\p^2\p_{\mu} 
c^a+d_nA_{\mu}^a\p_{\mu}c^a+\r.\nn\l. 
e_n\ep A_{\mu}^a\p_{\nu}A_{\mu}^b\p_{\nu}c^c 
+f_n  \ep A_{\mu}^a\p_{\mu}A_{\nu}^b\p_{\nu}c^c
+\r.\nn\l. g_nA_{\mu}^aA_{\mu}^a\p_{\nu}A_{\nu}^b c^b
+h_nA_{\mu}^aA_{\mu}^b\p_{\nu}A_{\nu}^a c^b
+\r.\nn\l. k_nA_{\mu}^aA_{\nu}^a\p_{\mu}A_{\nu}^b c^b
+j_nA_{\mu}^aA_{\nu}^b\p_{\mu}A_{\nu}^a c^b
+\r.\nn\l. l_nA_{\mu}^aA_{\nu}^b\p_{\mu}A_{\nu}^b c^a\]\ .\label{6,40}\eea
Thereby we see that this functional does not contain monomials of dimension 
lower than 3.

Inserting into (\ref{6,41}) (\ref{6,37}) and (\ref{6,40}) one finds, after 
some algebraic manipulations, the 
following constraints:
\bea b_n=-ga_n\quad ,\quad f_n+e_n+gc_n=0\nn
h_n=k_n=j_n\quad ,\quad l_n=2g_n\ ,\label{6,42}\eea
therefore the general solution of the perturbative consistency condition 
contains six free parameters.

Now, given
 the most general local approximant to $\S_n$ 
satisfying (\ref{6,30}) is:
\bea \Sl_n=
\int d^4x\ \[{z_1\o4} G^a_{\mu\nu} G^a_{\mu\nu} 
-z_2\(\c_{\mu}^a D_{\mu} c^a
+\zeta^a {g\epsilon^{abc}\o2}c^bc^c\)\r.\nn\l.+z_3\( D_{\mu}A_{\nu}^a  G^a_{\mu\nu}
+\c_{\mu}^a \p_{\mu} c^a\)
+x_1\ep\c_{\mu}^aA_{\mu}^bc^c +x_2 \p_{\mu} A_{\mu}^a\p_{\nu}A_{\nu}^a \r.\nn\l.
+x_3A_{\mu}^aA_{\mu}^a+ x_4\ep A_{\mu}^aA_{\nu}^b\p_{\mu}A_{\nu}^c
+x_5A_{\mu}^aA_{\mu}^bA_{\nu}^aA_{\nu}^b+x_6\(A_{\mu}^aA_{\mu}^a\)^2 
\].\label{6,43}\eea
The first three terms in the right-hand side are annihilated by ${\cal D}$; this is
 obviously true for the first term that depends only on the gauge field and is gauge 
invariant; it is also obvious for the second term that added to the first one gives 
the action (\ref{6,37}). The third term is obtained from the action 
introducing the scaling 
transformation: $A_{\mu}\rightarrow Z A_{\mu}\ ,\  \c_{\mu}\rightarrow Z^{-1}\c_{\mu}
$ and taking the $Z$-derivative for $Z=1$; it is annihilated by ${\cal D}$ since the
scaling transformation leaves the differential operator (\ref{6,34}) invariant.
It is a little lengthy but straightforward to verify that
$\hDl_n$ in (\ref{6,41}) with the constraints (\ref{6,42}) satisfies:
\be \hDl_n={\cal D}\bar\Sl_n\ ,\label{6,44}\ee
if the coefficients of $\bar\Sl_n$ are chosen according
\bea a_n=-\bar x_1 \quad ,\quad c_n=-2\bar x_2\nn
d_n=2\bar x_3 \quad ,\quad e_n=\bar x_4\nn
g_n=-\bar x_4-4\bar x_6\quad ,\quad h_n=\bar x_4-4\bar x_5\ .\label{6,45}\eea
Let us now consider the structure of $\hDl$ at the $n^{th}$ order in 
$\hbar$ as it can be computed from (\ref{6,33b}). This is written:
\bea \hDl_n=\nn
\IP\[\dea \Sl_0\deg\Sl_n+\deg \Sl_0\dea\Sl_n+\r.\nn\l. \dec 
\Sl_0\dez\Sl_n+  
\dez \Sl_0\dec\Sl_n+\r.\nn\l.\sum_{k=1}^{n-1}
{\cal T}_5\(\dea \S_k\deg\S_{n-k}+\dec \S_k\dez\S_{n-k}\)-\r.\nn\l.
 \hbar{\cal T}_5\kp
\(\dea\deg+\dec\dez\)\S_{n-1}\]\ .\label{6,46}\eea
On account of (\ref{6,44}),  (\ref{6,46}) can be written:
\bea {\cal D}\Sl_n +\nn\IP\[\sum_{k=1}^{n-1}
{\cal T}_5\(\dea \S_k\deg\S_{n-k}+\r.\r.\nn\l.\l.\dec \S_k\dez\S_{n-k}\)-
\r.\nn\l.
 \hbar{\cal T}_5\kp
\(\dea\deg+\dec\dez\)\S_{n-1}\]\equiv\nn
{\cal D}\Sl_n +\Omega_n={\cal D}\bar\Sl_n\ ,\label{6,47}\eea
from which we see that there exists a 4-dimensional integrated local 
functional $\Sigma_n=\bar\Sl_n-\Sl_n$ such that 
\be \Omega_n={\cal D}\Sigma_n\ .\label{6,48}\ee
Taking into account this result and (\ref{6,46}) the $n^{th}$ order 
fine-tuning equation is written:
\be {\cal D}\( \Sl_n +\Sigma_n\)=0\ ,\ee and it is solved by
\be \Sl_n=-\Sigma_n\ .\ee
This proves the iterative solvability of the fine-tuning equation.

\salto
\sec{Further comments.}
\salto

As it is apparent from the above analysis the crucial step of this proof 
has been the fact that the consistency condition can be written in the 
form
\be {\cal D}\hDl_n=0\ ,\label{6,50}\ee 
and the result that the general solution of this equation is
\be \hDl_n={\cal D}\bar\Sl_n\ .\ee In the framework of a more general gauge 
theory one can always reduce the perturbative problem to the study of an 
equation completely analogous to (\ref{6,50}) and in particular to the 
comparison of the kernel of the nilpotent operator with its image. 
In this way the 
solvability of the fine-tuning equation is in general reduced to the 
triviality of a certain class of the so-called 
BRS cohomology. This is exactly what we 
have verified in the present example.

Of course, our proof is strongly dependent on the chosen perturbative 
framework. However the idea of making the low-energy vertices of the 
breaking to vanish by a fine tuning of the parameters of the theory could 
well work beyond perturbation theory.

The first question to answer in this direction is how this possibility is 
related to the linear consistency condition (\ref{6,35}). In particular it 
can be interesting to notice that, if we write (\ref{6,35}) in the form:
\be \hat{\cal D}\hDe=0\ ,\ee
the differential operator $\hat{\cal D}$ is not nilpotent. Indeed in the 
case of anticommuting $\c$ and commuting $\f$ one has ($\hbar=1$)
\be{\hat{\cal D}}^2=\int d^4p\[{\d\o\d\c (p)} \hDe{\d\o\d\f (-p)}-
{\d\o\d\f (-p)}\hDe{\d\o\d\c (p)} \]\ .\ee
Here we have assumed $\f$ commuting and $\c$ anticommuting. 

Thus the 
nilpotency of the perturbative coboundary operator ${\cal D}$ is violated 
beyond the pertubative level by terms of order $\hDe$.
\vfill\eject

\end{document}